\documentstyle[12pt,epsfig]{article}
\def\be{\begin{equation}}
\def\ee{\end{equation}}
\catcode `\@=11 \@addtoreset{equation}{section}

\newcommand{\plot}[1]
{\begin{center} \epsfxsize=8cm
\parbox{\epsfxsize}{\epsffile{#1}}
\end{center}}

\begin{document}
\begin{titlepage}
\title{\Large\bf Toroidal configuration of the orbit of the electron of
the hydrogen atom under strong external magnetic fields}
\author{\normalsize\bf A.K. Aringazin$^{1,2}$}

\date{\normalsize
$^1$Department of Theoretical Physics, Karaganda State University,
Karaganda 470074 Kazakstan\\
ascar@ibr.kargu.krg.kz\\
$^2$Institute for Basic Research, P.O. Box 1577, Palm Harbor, FL 34682, USA\\
ibr@gte.net\\[1cm]
May 10, 2001}

\maketitle

\abstract{In this paper we overview some results on the hydrogen atom
in external static uniform magnetic fields. We focus on the case of
very strong magnetic field, $B \gg B_0=2.3\cdot 10^{9}$ Gauss, use
various approximate models and, particularly, in the adiabatic
approximation have calculated exactly the integral defining the
effective potential. This potential appears to be finite at $z=0$. Our
consideration of the problem of highly magnetized atoms and molecules
is motivated by the recently developed MagneGas technology by Santilli
(http://www.magnegas.com). The ground state electron charge
distribution of the hydrogen atom in an intense magnetic field is of a
toroidal form, in agreement with that studied by Santilli. This
physical picture is at the foundation of the new chemical species of
magnecules proposed by Santilli.}
\end{titlepage}

\section{Introduction}

Weak external static uniform magnetic field $B$ causes anomalous Zeeman
splitting of the energy levels of the hydrogen atom, with ignorably
small effect on the charge distribution of the electron. In the case of
a more intense magnetic field which is strong enough to cause
decoupling of a spin-orbital interaction (in atoms), $e\hbar B/2mc >
\Delta E_{jj'} \simeq 10^{-3}$ eV, i.e. $B \simeq 10^5$~Gauss, a normal
Zeeman effect is observed, again with ignorably small deformation of
the electron orbits.

In the case of {\sl weak} external magnetic field $B$, one can ignore
the quad\-ra\-tic term in the field $B$ because its contribution is
small in comparison with that of the other terms in Schr\"odinger
equation so that the {\sl linear} approximation in the field $B$ can be
used. In such a linear approximation, the wave function of electron
remains unperturbed, with the only effect being the well known Zeeman
splitting of the energy levels of the hydrogen atom. In both Zeeman
effects, the energy of interaction of electron with the magnetic field
is assumed to be much smaller than the binding energy of the hydrogen
atom, $e\hbar B/2mc \ll me^4/2\hbar^2 = 13.6$ eV, i.e. the intensity of
the magnetic field is much smaller than some characteristic value, $B
\ll B_0 = 2.4\cdot 10^{9}$ Gauss = 240000 Tesla (1 Tesla = $10^4$
Gauss). Thus, the action of a weak magnetic field can be treated as a
small perturbation of the hydrogen atom.

In the case of \textsl{very strong} magnetic field, $B \gg B_0$, the
quadratic term in the field $B$ makes a great contribution and can not
be ignored. Calculations show that a considerable deformation of the
electron charge distribution in the hydrogen atom occurs. Namely, under
the influence of a very strong external magnetic field a magnetic
confinement takes place, i.e. in the plane perpendicular to the
direction of magnetic field the electron dynamics is determined mainly
by the action of the magnetic field, while the Coulomb interaction of
the electron with the nucleus can be viewed as a small perturbation.
This adiabatic approximation allows one to separate variables in
Schr\"odinger equation \cite{Sokolov}. At the same time, in the
direction along the direction of the magnetic field the motion of
electron is governed both by the magnetic field effects and the Coulomb
interaction of the electron with the nucleus.

In this paper we briefly review some results on the hydrogen atom in
very strong external static uniform magnetic fields, focusing on the
basic physical picture derived from the Schr\"odinger equation. Our
consideration of the problem of atoms and molecules exposed to strong
external magnetic field is related to MagneGas technology and
PlasmaArcFlow hadronic molecular reactors recently developed by
Santilli \cite{Santilli}.

The highest intensities maintained macroscopically at large distances
in modern magnet laboratories are of the order of
$10^{5}...10^{6}$~Gauss ($\sim$ 50 Tesla), i.e. much below
$B_0=2.4\cdot 10^9$ Gauss ($\sim 10^5$ Tesla). An extremely intense
external magnetic field, $B \ge B_S=B_0/\alpha^2= 4.4\cdot
10^{13}$~Gauss, corresponds to the interaction energy of the order of
mass of electron, $mc^2=0.5$ MeV; $\alpha=e^2/\hbar c$ is the fine
structure constant. In this case, despite the fact that the extremely
strong magnetic field does not make vacuum unstable in respect to
creation of electron-positron pairs, one should account for
relativistic and quantum electrodynamics (QED) effects, and invoke
Dirac or Bethe-Salpeter equation. Such intensities are not currently
available in laboratories. However, these are of interest in
astrophysics, for example in studying an atmosphere of neutron stars
and white dwarfs which is characterized by $B \simeq 10^{9}\dots
10^{13}$ Gauss.

In this paper, we shall not consider intensities as high as Schwinger
value $B_S$, and restrict our review by the intensities $2.4\cdot
10^{10} \leq B \leq 2.4\cdot 10^{11}$ Gauss ($10B_0 \dots 100B_0$), at
which (nonrelativistic) Schr\"odinger equation can be used to a very
good accuracy, and the adiabatic approximation can be made.

Relativistic and QED effects (loop contributions), as well as the
effects such as those related to the finite mass, size, and magnetic
moment of the nucleus, and the finite electromagnetic radius of
electron, reveal themselves even at low magnetic field intensities, and
can be accounted for as very small perturbations. These are beyond the
scope of the present paper, while being of much importance in the high
precision studies, such as those on stringent tests of the Lamb shift.

It should be noted that a locally high-intensity magnetic field may
arise in the plasma as the result of nonlinear effects, which can lead
to creation of stable self-confined structures having a nontrivial
topology with knots \cite{Faddeev}. Particularly, Faddeev and Niemi
\cite{Faddeev} recently argued that the static equilibrium
configurations within the plasma are topologically stable solitons,
that describe knotted and linked fluxtubes of helical magnetic fields.
In the region close to such fluxtubes, we suppose the magnetic field
intensity may be as high as $B_0$. In view of this, study of the action
of strong magnetic field and the fluxtubes of magnetic fields on atoms
and molecules becomes of much interest in theoretical and applicational
\textsl{plasmachemistry}.

Possible applications are in MagneGas technology, theoretical
foundations and exciting implications of which are developed by
Santilli \cite{Santilli}. We refer the reader to Ref.~\cite{Santilli}
for the recent studies, description, and applications of MagneGas
technology and PlasmaArcFlow reactors. Remarkable experimental results,
including Gas-Chromatographic Mass-Spectroscopic and InfraRed spectrum
data of Santilli's magnegas, clearly indicates the presence of
unconventional chemical species of {\it magnecules}, which have not
been identified as conventional molecules, and are supposed to be due
to the specific bonding of a magnetic origin. The basic physical
picture of the magnecules proposed by Santilli \cite{Santilli} is
related to the toroidal orbitals of the electrons of atoms exposed to a
very strong magnetic field. The results of the present paper
demonstrate this physical picture on the basis of the Schr\"odinger
equation.

As the result of the action of very strong magnetic field, atoms attain
great binding energy as compared to the case of zero magnetic field.
Even at intermediate $B\simeq B_0$, the binding energy of atoms greatly
deviates from that of zero-field case, and even lower field intensities
may essentially affect chemical properties of molecules of heavy atoms.
This enables creation of various other bound states in molecules,
clusters and bulk matter \cite{Sokolov, Santilli, Lai}.

The paper by Lai \cite{Lai}, who focused on very strong magnetic
fields, $B \gg B_0$, motivated by the astrophysical applications, gives
a good survey of the early and recent studies in this field, including
those on the intermediate range, $B \simeq B_0$, multi-electron atoms,
and H$_2$ molecule. Much number of papers using variational/numerical
and/or analytical approaches to the problem of light and heavy atoms,
ions, and H$_2$ molecule in strong magnetic field, have been published
within the last six years (see, e.g., references in \cite{Lai}).
However, highly magnetized molecules of heavy atoms have not been
systematically investigated. One of the surprising implications is that
for some diatomic molecules of heavy atoms, the molecular binding
energy is predicted to be several times bigger than the ground state
energy of individual atoms. \cite{Johnsen}

\section{Landau levels of a single electron}

To estimate intensity of the magnetic field which causes a considerable
deformation of the ground state electron orbit of the hydrogen atom,
one can formally compare the Bohr radius of the hydrogen atom in the
ground state, in zero external magnetic field, $a_0=\hbar^2/me^2 \simeq
0.53\cdot 10^{-8}$ cm = 1 a.u., with the radius of orbit of a single
electron moving in the external static uniform magnetic field $\vec B$.
Below, we give a brief summary of the latter issue.

\begin{figure}[ht]
\begin{center}
\epsfxsize=3cm
\parbox{\epsfxsize}{\epsffile{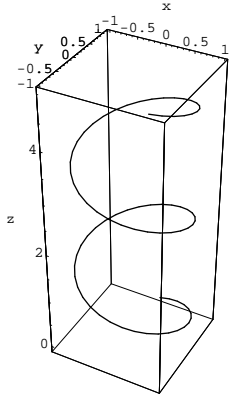}}
\end{center}
\caption{Classical orbit of the electron in an external static uniform
magnetic field $\vec B = (0,0,B)$ pointed along the $z$ axis. The
electron experiences a circular motion in the $x$, $y$ plane and a
linear motion in the $z$ direction (a helical curve).}
\label{Fig:class}
\end{figure}

The mean radius of the orbital of a single electron moving in a static
uniform magnetic field can be calculated exactly using Schr\"odinger
equation, and is given by
\be\label{LandauR}
R_n = \sqrt{\frac{n+1/2}{\gamma}},
\ee
where we have denoted
\be\label{gamma}
\gamma = \frac{eH}{2\hbar c},
\ee
$B$ is the intensity of the magnetic field pointed along the $z$ axis,
$\vec B = (0,0,B)$, $\vec r = (r,\varphi,z)$ in cylindrical
coordinates, and $n=0,1,\dots$ is the principal quantum number. Thus,
the radius of the orbit takes \textsl{discrete} set of values
(\ref{LandauR}), and is referred to as Landau radius. This is in
contrast to the well known \textsl{classical} motion of electron in the
external magnetic field (a helical curve is shown in
Fig.~\ref{Fig:class}), with the radius of the orbit being of a
continuous set of values. The classical approach evidently can not be
applied to study dynamics of the electron at atomic distances.

Corresponding energy levels $E_n$ of a single electron moving in the
external magnetic field are referred to as Landau energy levels,
\be\label{LandauE}
E_n = E_n^{\perp} + E_{k_z}^{\|} =
\hbar\Omega(n+\frac{1}{2}) + \frac{\hbar^2k^2_z}{2m},
\ee
where
\be\label{Omega}
\Omega = \frac{eH}{mc}
\ee
is so called cyclotron frequency, and $\hbar k_z$ is a projection of
the electron's momentum $\hbar\vec k$ on the direction of the magnetic
field, $-\infty <k_z<\infty$, $m$ is the mass of electron, and $-e$ is
the charge of electron.

Landau energy levels $E_n^{\perp}$ correspond to a discrete set of
round orbits of electron which are projected to the transverse plane.
The energy $E_{k_z}^{\|}$ corresponds to the free motion of electron in
parallel to the magnetic field (\textsl{continuous} spectrum), with a
conserved momentum $\hbar k_z$ along the magnetic field.

Regarding the above presented review of Landau's results, we remind
that in the general case of \textsl{uniform} external magnetic field
the coordinate and spin components of the total wave function of the
electron can always be separated.

The corresponding coordinate component of the total wave function of
the electron, obtained as an exact solution of Schr\"odinger equation
for a single electron moving in the external magnetic field with
vector-potential chosen as $A_r=A_z=0$, $A_\varphi = rH/2$),
\be\label{Schr0}
-\frac{\hbar^2}{2m}
\left(
\partial_r^2+\frac{1}{r}\partial_r+\frac{1}{r^2}\partial_\varphi^2
+\partial_z^2 -\gamma^2r^2 + 2i\gamma\partial_\varphi \right) \psi =
E\psi,
\ee
is of the following form \cite{Sokolov}:
\be\label{LandauPsi}
\psi_{n,s,k_z}(r,\varphi,z) =
\sqrt{2\gamma}I_{ns}(\gamma r^2)
\frac{e^{il\varphi}}{\sqrt{2\pi}}
\frac{e^{ik_zz}}{\sqrt{L}},
\ee
where $I_{ns}(\rho)$ is Laguerre function,
\be
I_{ns}(\rho) = \frac{1}{\sqrt{n!s!}}e^{-\rho/2}\rho^{(n-s)/2}Q^{n-s}_s(\rho);
\ee
$Q^{n-s}_s$ is Laguerre polynomial, $L$ is normalization constant,
$l=0, \pm 1, \pm 2, \dots$ is azimuthal quantum number, $s=n-l$ is the
radial quantum number, and we have denoted, for brevity,
\be
\rho = \gamma r^2.
\ee
The wave function (\ref{LandauPsi}) depends on the angle $\varphi$ as
$\exp[il\varphi]$, and the energy (\ref{LandauE}) does not depend on
the azimuthal quantum number $l$, so the system reveals a {\sl
rotational} symmetry in respect to the angle $\varphi$ and all the
Landau orbits are round.

Also, the energy (\ref{LandauE}) does not depend on the radial quantum
number $s$. This means a degeneracy of the system in respect to the
position of the center of the orbit, and corresponds to an "unfixed"
position of the center. At $n-s>0$, the origin of the coordinate system
is inside the orbit while at $n-s<0$ it is outside the orbit. The mean
distance $d$ between the origin of coordinate system and the center of
orbit takes a discrete set of values, and is related to the quantum
number $s$ as follows:
\be\label{d}
d = \sqrt{\frac{s+1/2}{\gamma}}.
\ee
Clearly, the energy (\ref{LandauE}) does not depend on this distance
because of the homogeneity of the magnetic field, so it does not depend
on $s$. However, it should be noted that the whole Landau orbit can not
move on the ($r, \varphi$) plane continuously along the variation of
the coordinate $r$. Instead, it can only "jump" to a certain distance,
which depends on the intensity of the magnetic field, in accord to
Eq.~(\ref{d}).

The spin components of the total wave function are trivially
\be\label{SpinPsi}
\psi(\frac{1}{2}) =
\left(\begin{array}{c}1\\0\end{array}\right), \quad
\psi(-\frac{1}{2}) =
\left(\begin{array}{c}0\\1\end{array}\right),
\ee
with the corresponding energies
\be\label{LandauS}
E_{spin}=\pm \mu_0 B,
\ee
to be added to the energy (\ref{LandauE}). Here,
\be
\mu_0 = \frac{e\hbar}{2mc} \simeq 9.3 \cdot 10^{-21} \mbox{
erg}\cdot\mbox{Gauss}^{-1} = 5.8 \cdot 10^{-9} \mbox{
eV}\cdot\mbox{Gauss}^{-1}
\ee
is Bohr magneton. Below, we do not consider the energy (\ref{LandauS})
related to the interaction of the spin of electron with the external
magnetic field because it decouples from the orbital motion and can be
accounted for separately. As to the numerical estimation, at $B=B_0=
2.4\cdot 10^{9}$ Gauss, the energy $E_{spin} \simeq \pm 13.6$ eV.

For the \textsl{ground} level, i.e. at $n=0$ and $s=0$, and zero
momentum of the electron in the $z$-direction, i.e. $\hbar k_z =0$, we
have from (\ref{LandauE})
\be\label{LandauE0}
E_0^{\perp} = \frac{e\hbar B}{2mc},
\ee
and due to Eq.~(\ref{LandauPsi}) the corresponding normalized ground
state wave function is
\be\label{psi000}
\psi_{000}(r,\varphi,z) = \psi_{000}(r) = \sqrt{\frac{\gamma}{\pi}}\,
e^{-\gamma r^2/2},
\ee
$\int_{0}^{\infty}\!\!\int_{0}^{2\pi}r dr d\varphi\ |\psi_{000}|^2 =1$.

\begin{figure}[ht]
\plot{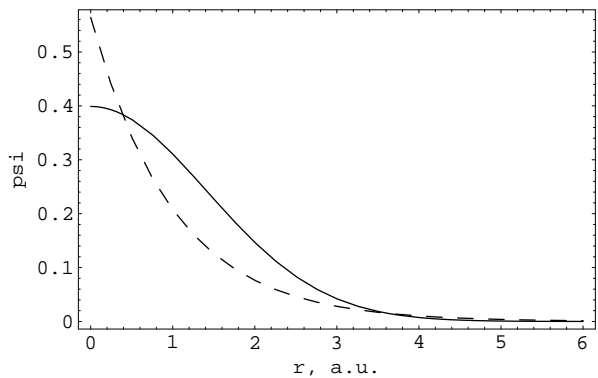} \caption{Landau ground state wave function of a
single electron, $\psi_{000}$ (solid curve), Eq. (\ref{psi000}), in
strong external magnetic field $B=B_0=2.4\cdot 10^{9}$ Gauss, as a
function of the distance $r$ in cylindrical coordinates, and (for a
comparison) the hydrogen ground state wave function (at \textsl{zero}
external magnetic field), $(1/\sqrt{\pi})e^{-r/a_0}$ (dashed curve), as
a function of the distance $r$ in spherical coordinates. The associated
probability densities are shown in Fig.~\ref{Fig:density}; 1 a.u. =
$a_0$ = $0.53\cdot 10^{-8}$ cm.} \label{Fig:psi000}
\end{figure}

\begin{figure}[ht]
\plot{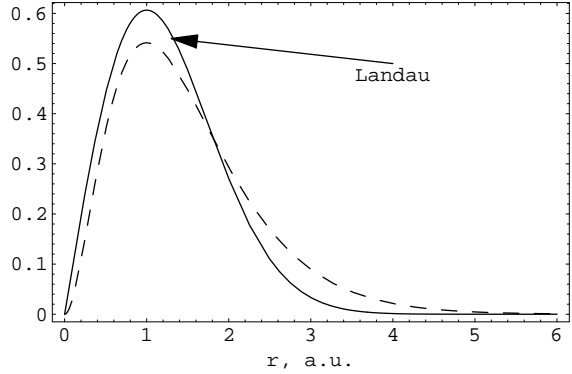} \caption{The probability density for the case of the
Landau ground state of a single electron, $2\pi r|\psi_{000}|^2$ (solid
curve), Eq. (\ref{psi000}), in the strong external magnetic field
$B=B_0=2.4\cdot 10^{9}$ Gauss, as a function of the distance $r$ in
cylindrical coordinates, and (for a comparison) probability density of
the hydrogen atom ground state (at \textsl{zero} external magnetic
field), $4\pi r^2|(1/\sqrt{\pi})e^{-r/a_0}|^2$ (dashed curve), as a
function of the distance $r$ in spherical coordinates. The associated
wave functions are shown in Fig.~\ref{Fig:psi000}; 1 a.u. = $0.53\cdot
10^{-8}$ cm.} \label{Fig:density}
\end{figure}

\begin{figure}[ht]
\plot{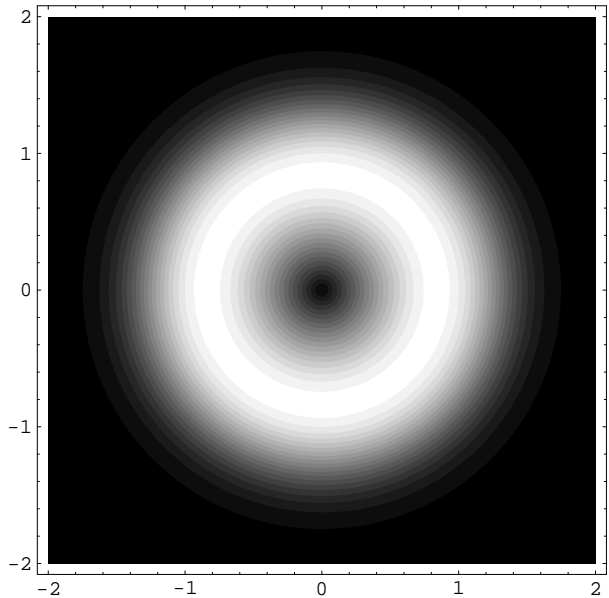} \caption{A contour plot of the $(r,\varphi)$
probability density for the case of the Landau ground state of a single
electron, $2\pi r|\psi_{000}|^2$, Eq. (\ref{psi000}), in the strong
external magnetic field $B=B_0=2.4\cdot 10^{9}$ Gauss, as a function of
the distance in a.u. (1 a.u. = $0.53\cdot 10^{-8}$ cm). Lighter area
corresponds to higher probability to find electron. The set of maximal
values of the probability density is referred to as an "orbit".}
\label{Fig:contour}
\end{figure}

The corresponding (smallest) Landau radius of the orbit of electron is
\be
R_0 = \sqrt{\frac{\hbar c}{e B}} \equiv \sqrt{\frac{1}{2\gamma}},
\ee
in terms of which $\psi_{000}$ reads
\be
\psi_{000} = \sqrt{\frac{1}{2\pi R_0^2}}\ e^{-\frac{r^2}{4R_0^2}}.
\ee

Figure~\ref{Fig:psi000} depicts ground state wave function of a single
electron, $\psi_{000}$, in strong external magnetic field
$B=B_0=2.4\cdot 10^{9}$ Gauss ($R_0 = 1$ a.u.), and (for a comparison)
of the hydrogen ground state wave function, at \textsl{zero} external
magnetic field, $(1/\sqrt{\pi})e^{-r/a_0}$. Figures~\ref{Fig:density}
and \ref{Fig:contour} display the associated probability density of
electron as a function of the distance $r$ from the center of the
orbit, the radius of which is about 1 a.u.

The condition that the Landau radius $R_0$ is smaller than the Bohr
radius, $R_0 < a_0$, which is adopted here as the condition of a
considerable "deformation" of the electron orbit of the hydrogen atom,
then implies
\be\label{Hcrit}
B > B_0= \frac{m^2ce^3}{\hbar^3} = 2.351\cdot 10^9 \mbox{ Gauss},
\ee
where $m$ is mass of electron. Equivalently, this deformation condition
corresponds to the case when the binding energy of the hydrogen atom,
$|E_0^{Bohr}|=|-me^4/2\hbar^2| = 0.5$ a.u. = 13.6 eV, is smaller than
the ground Landau energy $E_0^{\perp}$.

The above critical value of the magnetic field, $B_0$, is naturally
taken as an \textsl{atomic unit} for the strength of the magnetic
field, and corresponds to the case when the pure Coulomb interaction
energy of electron with nucleus is equal to the interaction energy of a
single electron with the external magnetic field,
$|E_0^{Bohr}|=E_0^{\perp}=13.6$ eV, or equivalently, when the Bohr
radius is equal to the Landau radius, $a_0=R_0=0.53\cdot 10^{-8}$ cm.

It should be stressed here that we take the characteristic parameters,
Bohr energy $|E_0^{Bohr}|$ and Bohr radius $a_0$, of the hydrogen atom
with a single purpose to establish criterium for the critical strength
of the external magnetic field, for the hydrogen atom under
consideration. For other atoms the critical value of the magnetic field
may be evidently different.

\section{Hydrogen atom in magnetic fields}

After we have considered quantum dynamics of a single electron in the
external magnetic field, we turn to consideration of the hydrogen atom
in the external static uniform magnetic field.

We choose cylindrical coordinate system $(r,\varphi,z)$, in which the
external magnetic field is $\vec B = (0,0,B)$, i.e., the magnetic field
is directed along the $z$-axis. General Schr\"odinger equation for the
electron moving around a fixed proton (Born-Oppenheimer approximation)
in the presence of the external magnetic field is
\be\label{Schr}
-\frac{\hbar^2}{2m} \left(
\partial_r^2+\frac{1}{r}\partial_r+\frac{1}{r^2}\partial_\varphi^2
+\partial_z^2 +\frac{2me^2}{\hbar^2\sqrt{r^2+z^2}} -\gamma^2r^2 +
2i\gamma\partial_\varphi \right) \psi = E\psi,
\ee
where $\gamma = eH/2\hbar c$. Note the presence of the term
$2i\gamma\partial_\varphi$, which is \textsl{linear} in $B$, and of the
term $-\gamma^2r^2$, which is \textsl{quadratic} in $B$. The other
terms do not depend on the magnetic field $B$.

The main problem in the nonrelativistic study of the hydrogen atom in
the external magnetic field is to solve the above Schr\"odinger
equation and find the energy spectrum.

In this equation, we can not directly separate {\sl all} the variables,
$r$, $\varphi$, and $z$, because of the presence of the Coulomb
potential, $e^2/\sqrt{r^2+z^2}$, which does not allow us to make a
direct separation in variables $r$ and $z$.

Also, for the general case of an arbitrary intensity of the magnetic
field, we can not ignore the term $\gamma^2r^2$ in Eq. (\ref{Schr}),
which is quadratic in $B$, since this term is not small at
\textsl{high} intensities of the magnetic field.

For example, at small intensities, $B \simeq 2.4\cdot 10^4$ Gauss $=
2.4$ Tesla, the parameter $\gamma = eH/2\hbar c \simeq 1.7\cdot
10^{11}$ cm$^{-2}$ so that for $\langle r\rangle \simeq 0.5\cdot
10^{-8}$ cm we get the estimation $\gamma^2\langle r^2\rangle \simeq
10^{6}$ cm$^{-2}$ and Landau energy $\hbar\Omega/2$ is of the order of
$10^{-4}$ eV, while at high intensities, $B \simeq B_0 = 2.4\cdot
10^{9}$ Gauss $=2.4\cdot 10^5$ Tesla, the parameter $\gamma=\gamma_0
\simeq 1.7\cdot 10^{16}$ cm$^{-2}$ so that the estimations are:
$\gamma^2\langle r^2\rangle \simeq 10^{16}$ cm$^{-2}$ (greater by ten
orders) and the ground Landau energy is about 13.6 eV (greater by five
orders).

Below, we turn to approximate \textsl{ground state} solution of
Eq.~(\ref{Schr}) for the case of \textsl{very strong} magnetic field.

\subsection{Very strong magnetic field}\label{Sec:Strong}

Let us consider the approximation of a {\sl very} strong magnetic
field,
\be
B \gg B_0 = 2.4\cdot 10^{9} \mbox{ Gauss}.
\ee
Under the above condition, in the transverse plane the Coulomb
interaction of the electron with the nucleus is not important in
comparison with the interaction of the electron with the external
magnetic field. So, in accord to the exact solution (\ref{LandauPsi})
for a single electron, one can seek for an approximate ground state
solution of Eq. (\ref{Schr}) in the form (see, e.g. \cite{Sokolov}) of
factorized transverse and longitudinal parts,
\be\label{psi00}
\psi = e^{-\gamma r^2/2}\chi(z),
\ee
where we have used Landau wave function (\ref{psi000}), $s=0$, $l=0$,
and $\chi(z)$ denotes the longitudinal wave function to be found. This
is so called \textsl{adiabatic approximation}. The charge distribution
of the electron in the $(r, \varphi)$ plane is thus characterized by
the Landau wave function $\sim e^{-\gamma r^2/2}$, i.e. by the
azimuthal symmetry.

In general, the adiabatic approximation corresponds to the case when
the transverse motion of electron is totally determined by the intense
magnetic field, which makes it "dance" at its cyclotron frequency.
Specifically, the radius of the orbit is then \textsl{much smaller}
than the Bohr radius, $R_0 \ll a_0$, or, equivalently, the Landau
energy of electron is much bigger than the Bohr energy $E_0^{\perp} \gg
|E_0^{Bohr}|= 13.6$ eV. In other words, this approximation means that
the interaction of electron with the nucleus in the transverse plane is
ignorably small (it is estimated to make about 2\% correction at
$B\simeq 10^{12}$ Gauss), and the energy spectrum in this plane is
defined solely by the Landau levels. The remaining problem is thus to
find the longitudinal energy spectrum, in the $z$ direction.

Inserting wave function (\ref{psi00}) into the Schr\"odinger equation
(\ref{Schr}), multiplying it by $\psi^*$, and integrating over
variables $r$ and $\varphi$ in cylindrical coordinate system, we get
the following equation characterizing the $z$ dependence of the wave
function:
\be\label{Eqz}
\left(-\frac{\hbar^2}{2m}\frac{d^2}{dz^2} + \frac{\hbar^2\gamma}{m} +
C(z)\right)\chi(z) = E \chi(z),
\ee
where
\be\label{C}
C(z)= - \sqrt{\gamma}\, e^2 \int_{0}^{\infty}\!
\frac{e^{-\rho}}{\sqrt{\rho +\gamma z^2 }}\, d\rho = -
e^2\sqrt{\pi\gamma}\, e^{\gamma z^2}\mathrm{erfc}(\sqrt{\gamma}|z|),
\ee
erfc$(x) = 1-$erf$(x)$, and
\be
\mathrm{erf}(x) = \frac{2}{\sqrt{\pi}}\int\limits_{0}^{x}e^{-t^2}dt
\ee
is error function.

\begin{figure}[ht]
\plot{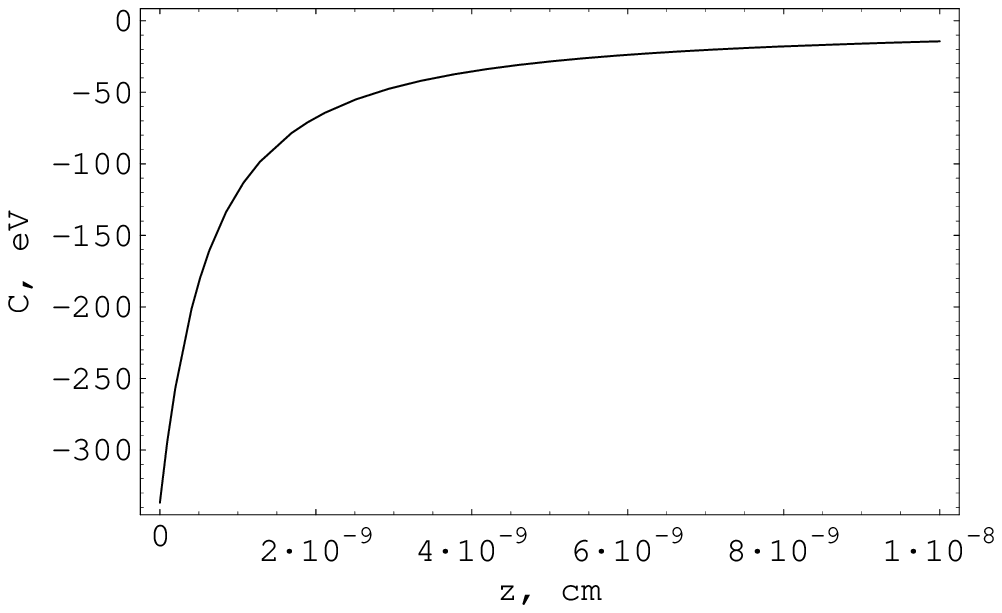} \plot{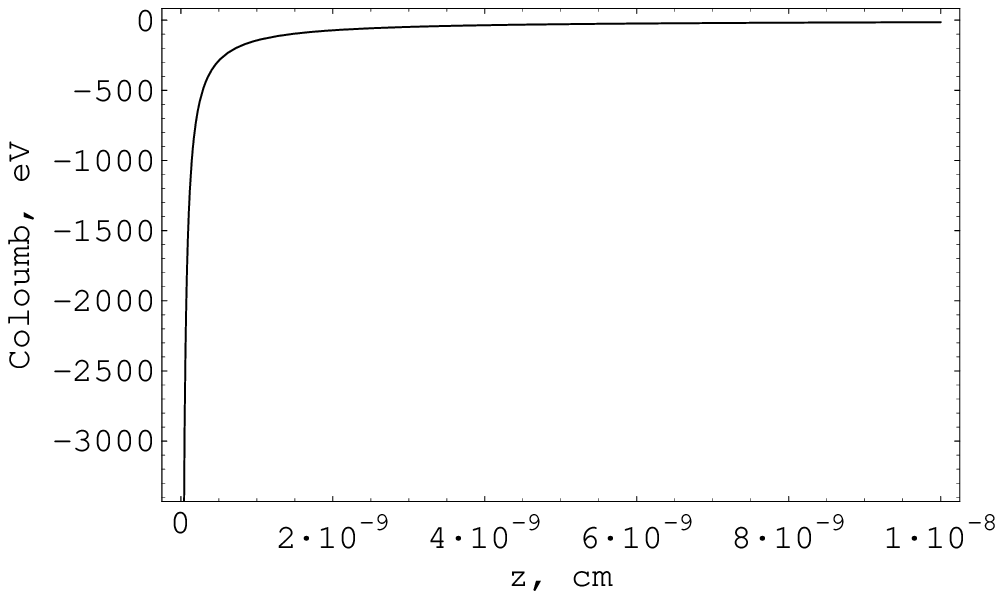} \caption{The exact effective potential
$C(z)$ (top panel), at $B=2.4\cdot 10^{11}$ Gauss $=100B_0$,
$C(0)=-337$ eV, and (for a comparison) Coulomb potential $-e^2/|z|$
(bottom panel).} \label{Fig:c}
\end{figure}

In Fig.~\ref{Fig:c}, the exact potential $C(z)$ is plotted, at
$B=2.4\cdot 10^{11}$ Gauss $=100B_0$, for which case the lowest point
of the potential is about $C(0)\simeq -337$ eV. In this case, in the
$(r,\varphi)$ plane, the hydrogen atom is characterized by the Landau
energy $E_0^{\perp}=1360$ eV $\gg 13.6$ eV, Landau radius $R_0=0.5\cdot
10^{-9}$ cm $\ll 0.5\cdot 10^{-8}$ cm, and the cyclotron frequency is
$\Omega=4\cdot 10^{18}$ sec$^{-1}$. For a comparison, the Coulomb
potential $-e^2/|z|$ is also shown. One can see that the Coulomb
potential does not reproduce the effective potential $C(z)$ to a good
accuracy at small $z$.

The arising effective potential $C(z)$ is of a nontrivial form, which
does not allow us to solve Eq. (\ref{Eqz}) analytically. Below we
approximate it by simple potentials, to make estimation on the ground
state energy and wave function of the hydrogen atom.

\subsubsection{The Coulomb potential approximation}\label{Cappr}

At high intensity of the magnetic field, $\gamma \gg 1$ so that under
the condition $\gamma \langle z^2\rangle \gg 1$ we can ignore $\rho$ in
the square root in the integrand in Eq. (\ref{C}). Then, one can
perform the simplified integral and obtain the result
\be\label{Ccoulomb}
C(z) \simeq V(z) = -\frac{e^2}{|z|}, \ \ \mbox{ at } \gamma \langle
z^2\rangle \gg 1,
\ee
which appears to be a pure Coulomb interaction of electron with the
nucleus, in the $z$ direction. Therefore, Eq. (\ref{Eqz}) reduces to
\textsl{one-dimensional} Schr\"odinger equation for Coulomb potential,
\be\label{1dim}
\left( \frac{\hbar^2}{2m}\frac{d^2}{dz^2} +\frac{e^2}{|z|} +
\frac{\hbar^2\gamma}{m} + E \right)\chi(z) = 0.
\ee
In atomic units ($e=\hbar = m=1$), using the notation
\be\label{E'}
E' = \frac{\hbar^2\gamma}{m} + E, \quad n^2 = \frac{1}{-2E'},
\ee
and introducing new variable
\be\label{varx}
x = \frac{2z}{n},
\ee
we rewrite the above equation in the following form:
\be
\left[\frac{d}{dx^2}+\left(-\frac{1}{4}+ \frac{n}{x}\right)\right]
\chi(x) =0,
\ee
where we assume, to simplify representation, $x>0$. Introducing new
function $v(x)$,
\be
\chi(x) = x e^{-x/2} v(x),
\ee
we get the final form of the equation,
\be
xv'' + (2-x)v' - (1-n)v =0.
\ee
 Note that this equation is identical to that obtained for the radial
part of the three-dimensional Schr\"odinger equation for the hydrogen
atom, at orbital quantum number zero. We note that it is a particular
case of Cummer's equation,
\be
xv'' + (b-x)v' - a v =0,
\ee
the general solution of which is given by
\be
v(x) = C_1\, {_1}\!F_{1}(a,b,x) + C_2 U(a,b,x)
\ee
where
\be
{_1}\!F_{1}(a,b,x) = \frac{\Gamma(b)}{\Gamma(b-a)\Gamma(a)}
\int_{0}^{1}\! e^{xt} t^{a-1}(1-t)^{b-a-1}\, dt
\ee
and
\be
U(a,b,x) = \frac{1}{\Gamma(a)} \int_{0}^{\infty}\! e^{-xt}
t^{a-1}(1+t)^{b-a-1}\, dt
\ee
are the confluent hypergeometric functions, and $C_{1,2}$ are
constants. In our case, $a=1-n$ and $b=2$.

For the wave function $\chi(x)$, we then have
\be\label{1solution}
\chi(x) = |x| e^{-|x|/2}\left[C_1^\pm {_1}\!F_{1}(1-n,2,|x|) + C_2^\pm
U(1-n,2,|x|) \right],
\ee
where $x$ given by Eq.~(\ref{varx}) is now allowed to have any real
value, and the "$\pm$" sign in $C_{1,2}^{\pm}$ corresponds to the positive
and negative values of $x$, respectively (the modulus sign is used for
brevity).

Now we turn to the normalization issue.

The first hypergeometric function ${_1}\!F_{1}(1-n, 2,x)$ is finite at
$x=0$ for any $n$. At big $x$, it diverges exponentially, unless $n$ is
an integer number, $n = 1,2, \dots,$ at which case it diverges
polynomially.

The second hypergeometric function $U(1-n,2,x)$ behaves differently,
somewhat as a mirror image of the first one. In the limit $x\to 0$, it
is finite for integer $n=1,2,3,\ldots$, and diverges as $1/x$ for
noninteger $n>1$ and for $0\leq n<1$. In the limit $x\to \infty$, it
diverges polynomially for integer $n$, tends to zero for noninteger
$n>1$ and for $n=0$, and diverges for noninteger $0<n<1$. In
Figs.~\ref{Fig:uf1} and \ref{Fig:uf2} of Appendix, the hypergeometric
functions ${_1}\!F_{1}(1-n, 2,x)$ and $U(1-n,2,x)$ for $n=0, 1/4, 1/2,
1, 3/2, 2, 5/2, 3$ are depicted ($n$ may be taken, for example,
$n=0.133$ as well).

In general, because of the prefactor $xe^{-x/2}$ in the solution
(\ref{1solution}) which cancels some of the divergencies arising from
the hypergeometric functions, we should take into account \textsl{both}
of the two linearly independent solutions, to get the most general form
of normalizable wave functions.

As a consequence, the eigenvalues may \textsl{differ} from those
corresponding to $n=1,2, \ldots$ (which is an analogue of the principal
quantum number in the ordinary hydrogen atom) so that $n$ is allowed to
take some \textsl{non-integer} values from 0 to $\infty$, provided that
the wave function is normalizable.

We are interested in the ground state solution, which is an even state.
For even states, in accord to the symmetry of the wave function under
the inversion $z\to-z$, we have
\be\label{evencond}
C_1^+ = C_1^-, \quad C_2^+ = C_2^-, \quad \chi'(0) = 0.
\ee
Also, since $n=1$ gives $E'=-1/(2n^2) = -1/2$ a.u., we should seek a
normalizable wave function for $n$ in the interval $0<n<1$, in order to
achieve the energy value, which is lower than $-1/2$ a.u. If it is
successful, the value $n=1$ indeed does not characterize the ground
state. Instead, it may correspond to some excited state.

One can see that the problem is in a remarkable difference from the
ordinary three-dimensional problem of the hydrogen atom, for which the
principal quantum number $n$ must be integer to get normalizable wave
functions, and the value $n=1$ corresponds to the lowest energy. The
distinction is due to the fact that the additional factor $x$ in the
solution (\ref{1solution}) does not cancel in the total wave
function\footnote{In the three-dimensional Coulomb problem it does
cancel in the total wave function.}.

We restrict consideration by the even state so we can focus on the $x
\ge 0$ domain, and drop the modulus sign,
\be
\chi(x) = xe^{-x/2}\left[C_1^{\pm}\, {_1}\!F_{1}(1-n, 2,x) + C_2^{\pm}
U(1-n,2,x)\right].
\ee

In summary, the problem is to identify a \textsl{minimal} value of $n$
in the interval $0<n \leq 1$, at which the wave function $\chi(x)$
represented by a combination of the \textsl{two} linearly independent
solutions is normalizable and obeys the conditions (\ref{evencond}).

The condition $\chi'(0)=0$ implies
\begin{equation}
\begin{array}{c}
\frac{1}{2}e^{-x/2}\left[C_1 \left( (2-x){_1}\!F_{1}(1-n, 2,x)+
(1-n)x{_1}\!F_{1}(2-n, 3,x)\right)+ \right.\\[10pt]
\left. + C_2\left( (2-x)U(1-n,2,x)- 2(1-n)x U(2-n,3,x)\right)
\right]_{|x=0}= 0.
\end{array}
\end{equation}
 The function ${_1}\!F_{1}(1-n,2,x)$ is well-defined at $x=0$ for any $n$.
The function $U(1-n,2,x)$ for small $x$ behaves as
$x^{-1}\Gamma^{-1}(1-n)$ so for $x=0$ it appears to be infinite, unless
$n=1$. The function $U(2-n,3,x)$ for small $x$ behaves as
$x^{-2}\Gamma^{-1}(2-n)$ so for $x=0$ it diverges ($\Gamma(a)$ is Euler
gamma function). Therefore, there is no way to satisfy the above
condition unless we put $C_2=0$, i.e. eliminate the second
hypergeometric function, $U(1-n,2,x)$, as an unphysical solution. As
the result, only the first hypergeometric function determines the wave
function, at which case it is well-defined only for integer $n$.

For the first hypergeometric function as the remaining solution, the
situation is well known,
\be
{_1}\!F_{1}(a,b,|x|)_{|x=0} = 1, \quad
 \frac{d}{dx}{_1}\!F_{1}(a,b,x)_{|x=0} = \frac{a}{b}.
\ee
We take $n=1$ as this value implies the lowest energy $E'$ in this
case. The energy for $n=1$ is $E' = -1/2$ a.u., i.e.
\be\label{Bohrz}
E= -\frac{me^4}{2\hbar^2} - \frac{1}{2}\hbar\Omega,
\ee
where we have used $\hbar^2\gamma/m = \frac{1}{2}\hbar\Omega$ due to
Eqs.~(\ref{gamma}) and (\ref{Omega}).

Thus, the particular state $n=1$ of the hydrogen atom is characterized
by usual Bohr distance $a_0$ in the $z$ direction, with the
longitudinal wave function
\be\label{chi}
\chi(z) \simeq |z| e^{-|z|/a_0}.
\ee
In addition, it is easy to check directly, without using the general
hypergeometric function technique, that $\chi(z) = ze^{-z}$ satisfies
the equation $(\frac{1}{2}d^2/dz^2+1/z+\epsilon)\chi(z) = 0$, for
$\epsilon= -1/2$.

In the $(r,\varphi)$ plane ($n=0, s=0$) the orbit is characterized by
the Landau radius $R_0\ll a_0$, with the Landau wave function $\simeq
e^{-r^2/4R_0^2}$, and the Landau energy $E_0^{\perp}$ given by the
second term in Eq.~(\ref{Bohrz}). Thus, the $n=1$ wave function in the
Coulomb approximation is written as
\be\label{psi000z}
\psi(r,\varphi,z) \simeq \sqrt{\frac{\gamma}{\pi}}\, ze^{-\gamma r^2/2
-|z|/a_0} = \sqrt{\frac{1}{2\pi R_0^2}}\ z e^{-\frac{r^2}{4R_0^2}-
\frac{|z|}{a_0}}.
\ee
Also, spin of the electron for the ground state is aligned antiparallel
to the magnetic field.

\begin{figure}[ht]
\begin{center} \epsfxsize=5.5cm
\parbox{\epsfxsize}{\epsffile{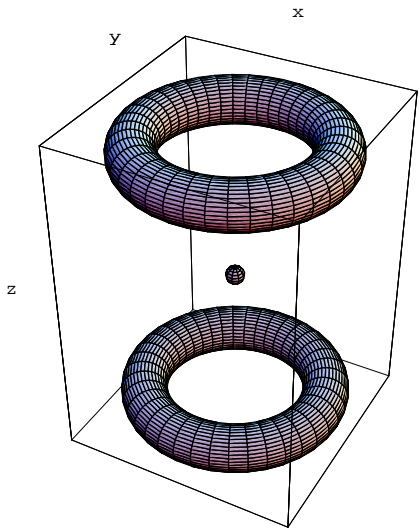}}
\end{center}
\caption{Schematic view on the hydrogen atom in the ground state, at
very strong external magnetic field $\vec B = (0,0,B)$, $B \gg
B_0=2.4\cdot 10^9$ Gauss, due to the Coulomb approximation approach,
Eq.~(\ref{psi000z}). One electron moves simultaneously on two Landau
orbits of radius $R_0$ which are shown schematically as torii in the
different ($x$, $y$) planes, one torus at the level $z=-a_0$ and the
other at the level $z=+a_0$, with the nucleus shown in the center at
$z=0$. Each torus represents the ($x$, $y$) probability distribution as
shown in Fig.~\ref{Fig:contour} but with small Landau radius, $R_0 \ll
a_0$. The spin of the electron is aligned antiparallel to the magnetic
field.} \label{Fig:double}
\end{figure}

The associated probability density is evidently of a cylindrical
(axial) symmetry and can be described as \textsl{two} Landau orbits of
radius $R_0$ in different $(r,\varphi)$ planes, one at the level
$z=-a_0$, and the other at the level $z=+a_0$, with the nucleus at
$z=0$, as schematically depicted in Fig.~\ref{Fig:double}. Presence of
two Landau orbits occurs in accord to the wave function
(\ref{psi000z}), which equals zero at $z=0$ and is symmetrical with
respect to the inversion, $z\to -z$. The electron moves simultaneously
on these two orbits.

The above electron charge distribution supports the study made by
Santilli \cite{Santilli} who proposed the polarized toroidal electron
orbit in the hydrogen atom under the action of strong magnetic field,
but the Coulomb approximation suggests the electron charge distribution
in the form of \textsl{two} identical coaxial round orbits separated by
relatively big vertical distance; see Fig.~\ref{Fig:double}.

Note that the size of the hydrogen atom in the $z$ direction is
predicted to be about 10 times bigger than that in the transverse
plane, i.e. $R_0 = 0.53\cdot 10^{-9}$ cm $ \simeq 0.1a_0$; the hydrogen
atom is thus highly elongated in the $z$ direction (this ratio is
\textsl{not} kept in Fig.~\ref{Fig:double}). The cyclotron frequency
(\ref{Omega}) is of about $4\cdot 10^{18}$ sec$^{-1}$. Classically, the
linear velocity of the electron on such an orbit is $v=2\pi R_0 \Omega$
and is estimated to be $v = 1.3\cdot 10^{10}$ cm/sec $\simeq 0.4c$.

We should to note here that Landau radius $R_0$, at $B=2.4\cdot
10^{11}$ Gauss $=100B_0$, is about ten times bigger than Compton
wavelength of electron, $\lambda_e/2\pi \simeq 0.4\cdot 10^{-10}$ cm,
hence we can ignore quantum electrodynamics effects.

Finally, it should be noted that the above used Coulomb approximation
is rather a crude one (see Fig.~\ref{Fig:c}). Indeed, the energy in the
$z$-direction does not depend on the magnetic field intensity while it
is obvious that it should depend on it. In the next Section, we
approximate the effective potential $C(z)$ with a better accuracy.

\subsubsection{The modified Coulomb potential approximation}\label{Cmodappr}

We note that due to the exact result (\ref{C}), the effective potential
$C(z)$ tends to zero as $z\to \infty$ (not a surprise). However, a
remarkable implication of the exact result is that $C(z)$ is {\sl
finite} at $z=0$, namely,
\be
C(0)=-\sqrt{\pi\gamma}\, e^2,
\ee
so that the effective potential $C(z)$ indeed can {\sl not} be well
approximated by the Coulomb potential, $-e^2/|z|$, at small $z$
($|z|<10^{-8}$ cm) despite the fact that at big values of $\gamma$
($\gamma\gg \gamma_0 \simeq 1.7\cdot 10^{16}$ cm$^{-2}$) the Coulomb
potential reproduces $C(z)$ to a good accuracy at long and medium
distances. Thus, the approximation considered in Sect.~\ref{Cappr}
appears to be not valid at (important) short distances.

Consequently, Eq.~(\ref{Eqz}) with the \textsl{exactly} calculated
$C(z)$ will yield the ground state of electron in the $z$ direction
which is drastically \textsl{different} from that with the Coulomb
potential considered in Sect.~\ref{Cappr}. Thus, very intense magnetic
field {\sl essentially affects dynamics of electron not only in the
$(r,\varphi)$ plane but also in the $z$ direction}. We expect that for
the case of exact potential (\ref{C}) the ground state is characterized
by {\sl much lower} energy in the $z$ direction as compared to that of
the one-dimensional Bohr state (\ref{Bohrz}).

The exact potential $C(z)$ can be well approximated by the
\textsl{modified} Coulomb potential,
\be\label{Cmod}
C(z) \simeq V(z) = -\frac{e^2}{|z|+z_0},
\ee
where $z_0$ is a parameter, $z_0\neq 0$, which depends on the field
intensity $B$ due to
\be\label{z0}
z_0 = - \frac{e^2}{C(0)} = \frac{1}{\sqrt{\pi\gamma}} =
\sqrt{\frac{2\hbar c}{\pi e B}}.
\ee
 The analytic advantage of this approximation is that $V(z)$ is
\textsl{finite} at $z=0$, being of Coulomb-type form.

Calculations for this approximate potential can be made in an
essentially the same way as in the preceding Section, with $|x|$ being
replaced by $|x|+x_0$, where $x_0>0$ (we remind that $x=2z/n$),
\be\label{1solutionx0}
\chi(x) = (|x|+x_0) e^{-(|x|+x_0)/2}\left[C_1^\pm
{_1}\!F_{1}(1\!-\!n,2,|x|\!+\!x_0) + C_2^\pm U(1\!-\!n,2,|x|\!+\!x_0)
\right].
\ee

An essential difference from the pure Coulomb potential case is that
the potential (\ref{Cmod}) has no singularities. The $x\to 0$
asymptotic of the associated second hypergeometric function,
$U(1-n,2,|x|+x_0)$ becomes well-defined for noninteger $n$. Thus, the
condition $\chi'(x)_{|x=0}=0$ does not lead to divergencies in the
second solution for noninteger $n<1$. It remains only to provide
well-defined behavior of the wave function at $x\to \infty$.

Analysis shows that \textsl{normalizable} wave functions, as a
combination of \textsl{two} linearly independent solutions, for the
modified Coulomb potential \textsl{does exist} for various
\textsl{non-integer}\footnote{This is the case when one obtains a kind
of "fractional" quantum number $n$ which does make sense.} $n$. We are
interested in the ground state solution, so we consider the values of
$n$ from 0 to 1. Remind that $E' = -1/(2n)^2$ so that for $n<1$ we
obtain the energy lower than $-0.5$ a.u.

For $n<1$, the first hypergeometric function is not suppressed by the
prefactor $x e^{-x/2}$ in the solution (\ref{1solutionx0}) at large $x$
so we are led to discard it as an unphysical solution by putting
$C_1=0$. A normalizable ground state wave function for $n<1$ is thus
may be given by the second term in the solution (\ref{1solutionx0}).
The condition $\chi'(x)_{|x=0}=0$ implies
\begin{equation}\label{Eq:cond}
\begin{array}{c}
\frac{1}{2}e^{-(x+x_0)/2}C_2[(2-x-x_0)U(1-n,2,x+x_0)-\qquad\qquad\\[5pt]
\qquad\qquad - 2(1-n)(x+x_0) U(2-n,3,x+x_0))]_{|x=0}= 0.
\end{array}
\end{equation}
 The l.h.s of this equation depends on $n$ and $x_0$, so we can select
some field intensity $B$, calculate associated value of the parameter
$x_0 = x_0(B)$ and find the value of $n$, from which we obtain the
ground state energy $E'$.

On the other hand, for the ground state this condition can be viewed,
{\it vice versa}, as an equation to find $x_0$ for some given value of
$n$ \cite{Heyl}. Hereby we reverse the order of the derivation, to
simplify numerical calculations.

For example, taking the noninteger value
\be
n=1/\sqrt{15.58} \simeq 0.253 < 1
\ee
we find from Eq. (\ref{Eq:cond}) numerically
\be
x_0 = 0.140841.
\ee
This value is in confirmation with the result $x_0=0.141$ obtained by
Heyl and Hernquist \cite{Heyl} (see Discussion). On the other hand,
$x_0$ is related in accord to Eq.~(\ref{z0}) to the intensity of the
magnetic field,
\be
x_0 = \frac{2z_0}{n} = \sqrt{\frac{8\hbar c}{\pi n^2 e B}},
\ee
from which we obtain $B \simeq 4.7 \cdot 10^{12}$ Gauss. Hence, at this
field intensity the ground state energy of the hydrogen atom is given
by $E' = -1/(2n^2) = - 15.58$ Rydberg.
\begin{figure}[ht]
\begin{center} \epsfxsize=7cm
\parbox{\epsfxsize}{\epsffile{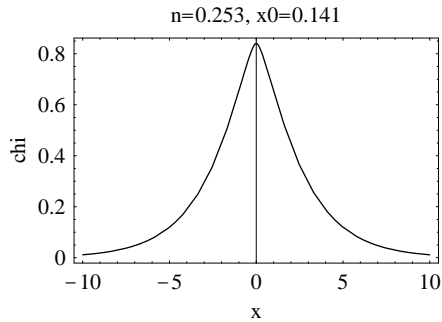}}
\end{center}
\caption{The longitudinal ground state wave function $\chi(x)$,
Eq.~(\ref{Eq:zground}), of the hydrogen atom in the magnetic field $B =
4.7\cdot 10^{12}$ Gauss; $n = 1/\sqrt{-2E}$, $x=2z/n$.}
\label{Fig:ground}
\end{figure}

\begin{figure}[ht]
\begin{center} \epsfxsize=7cm
\parbox{\epsfxsize}{\epsffile{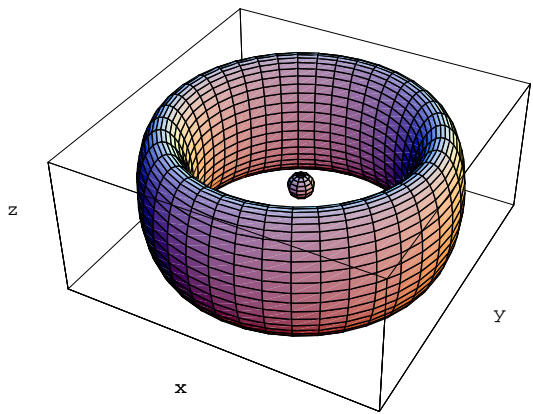}}
\end{center}
\caption{Schematic view of the hydrogen atom in the ground state, at
very strong external magnetic field $\vec B = (0,0,B)$, $B \gg
B_0=2.4\cdot 10^9$ Gauss, due to the \textsl{modified} Coulomb
approximation approach. The electron moves on the Landau orbit of small
radius $R_0 \ll 0.53\cdot 10^{-8}$ cm, which is shown schematically as
a torus. The vertical size of the atom is comparable to $R_0$.  Spin of
the electron is aligned antiparallel to the magnetic field.}
\label{Fig:single}
\end{figure}

The longitudinal ground state wave function is given by
\be\label{Eq:zground}
\chi(x) \simeq (|x|+x_0)e^{(|x|+x_0)/n} U(1-n, 2, |x|+x_0),
\ee
and is plotted in Fig.~\ref{Fig:ground}). The total wave function is
\be\label{Eq:ground}
\chi(x) \simeq \sqrt{\frac{1}{2\pi
R_0^2}}e^{-\frac{r^2}{4R_0^2}}(|x|+x_0)e^{(|x|+x_0)/n} U(1-n, 2,
|x|+x_0),
\ee
and the associated three-dimensional probability density is
schematically depicted in Fig.~\ref{Fig:single}.

In contrast to the double Landau-type orbit implied by the Coulomb
potential approximation, the modified Coulomb potential approach
provides qualitatively correct behavior and much better accuracy, and
suggests a \textsl{single} Landau-type orbit shown in
Fig.~\ref{Fig:single} for the \textsl{ground} state charge distribution
of the hydrogen atom. This is in full agreement with Santilli's study
\cite{Santilli} of the hydrogen atom in strong magnetic field.

Only excited states are characterized by the double Landau-type orbits
depicted in Fig.~\ref{Fig:double}. We shall not present calculations
for the excited states in the present paper, and refer the reader to
Discussion for some details on the results of the study made by Heyl
and Hernquist \cite{Heyl}.

\subsubsection{Variational and numerical solutions}

\begin{figure}[ht]
\begin{center} \epsfxsize=7cm
\parbox{\epsfxsize}{\epsffile{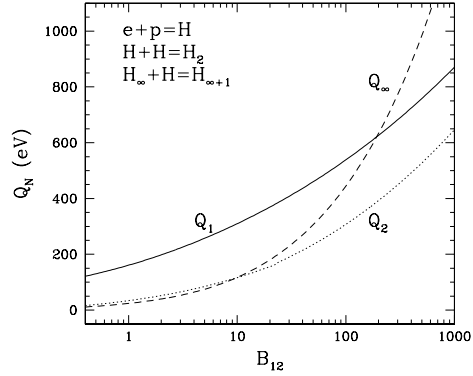}}
\end{center}
\vspace{-2cm} \caption{Ionization energies of the hydrogen atom and
H$_2$ molecule exposed to very strong magnetic field; $B_{12} = B\cdot
10^{-12}$ Gauss (reproduction of Fig.~1 by Lai~\cite{Lai}).}
\label{Fig:Lai1}
\end{figure}

Review of approximate, variational, and numerical solutions can be
found in the paper by Lai~\cite{Lai}. Accuracy of numerical solutions
is about 3\%, for the external magnetic field in the range from
$10^{11}$ to $10^{15}$ Gauss. Figure~\ref{Fig:Lai1} reproduces Fig.~1
of Ref.~\cite{Lai}. It represents the computed ionization energy $Q_1$
of the hydrogen atom and dissociation energy of H$_2$ molecule as
functions of the intensity of the magnetic field; $B_{12} = B\cdot
10^{-12}$ Gauss, i.e. $B_{12}=1$ means $B=10^{12}$ Gauss.

\begin{figure}[!hb]
\begin{center} \epsfxsize=7cm
\parbox{\epsfxsize}{\epsffile{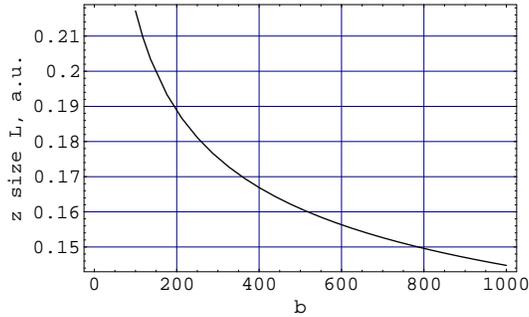}}
\end{center}
\caption{Longitudinal $z$-size of the hydrogen atom in the magnetic
field, Eq. (\ref{zsize}), as a function of the magnetic field
intensity; $b=B/B_0$, $B_0=2.4\cdot10^9$ Gauss, 1 a.u. =
$0.53\cdot10^{-8}$ cm.} \label{Fig:zsize}
\end{figure}

\begin{figure}[ht]
\begin{center} \epsfxsize=7cm
\parbox{\epsfxsize}{\epsffile{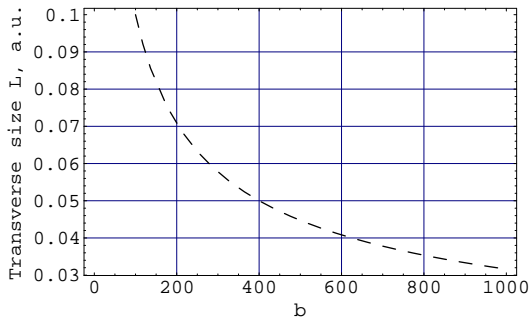}}
\end{center}
\caption{Transverse size of the hydrogen atom in the magnetic field,
Eq. (\ref{tsize}), as a function of the magnetic field intensity;
$b=B/B_0$, $B_0=2.4\cdot10^9$ Gauss, 1 a.u. = $0.53\cdot10^{-8}$ cm.}
\label{Fig:tsize}
\end{figure}

\begin{figure}[ht]
\begin{center} \epsfxsize=7cm
\parbox{\epsfxsize}{\epsffile{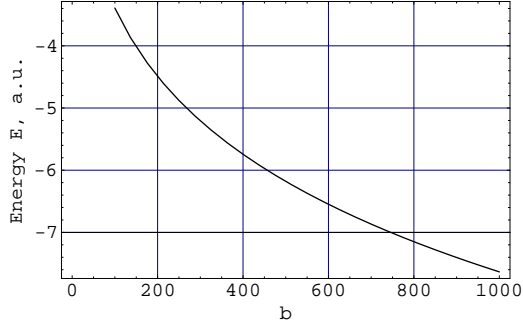}}
\end{center}
\caption{Variational ground state energy of the hydrogen atom in the
magnetic field, Eq. (\ref{energy}), as a function of the magnetic field
intensity; $b=B/B_0$, $B_0=2.4\cdot10^9$ Gauss, 1 a.u. = 27.2 eV.}
\label{Fig:energy}
\end{figure}

Due to the variational results \cite{Lai}, the $z$-size of the hydrogen
atom in the ground state is well approximated by the formula
(Fig.~\ref{Fig:zsize})
\be\label{zsize}
L_z \simeq \frac{1}{\ln(B/B_0)} \mbox{ a.u.};
\ee
the transverse (Landau) size is (Fig.~\ref{Fig:tsize})
\be\label{tsize}
L_{\perp} \simeq \frac{1}{\sqrt{B/B_0}}\mbox{ a.u.};
\ee
and the ground state energy is (Fig.~\ref{Fig:energy})
\be\label{energy}
E \simeq - 0.16 [\ln(B/B_0) ]^2 \mbox{ a.u.},
\ee
with the accuracy of few percents, for $b\equiv B/B_0$ in the range
from $10^2$ to $10^6$.

One can see for $B=100B_0$, that the variational study predicts the
ground state energy $E = - 3.4$ a.u. = $-92.5$ eV (much lower than the
value, $-0.5$ a.u., predicted by the $n=1$ Coulomb approximation in
Sect.~\ref{Cappr}), the transverse size $L_{\perp}$ of about 0.1 a.u.
(the same as in the Coulomb approximation) and the $z$-size $L_z$ of
about 0.22 a.u. (much smaller than the value, 1 a.u., predicted by the
Coulomb approximation). This confirms the result of the approximate
analytic approach, in which $n<1$ implies ground state energy much
lower than $-0.5$ a.u.

\section{Discussion}

In a physical context, the used adiabatic approximation implies that
the position of the nucleus is not "fixed" in the $(r,\varphi)$ plane.
Indeed, there is no reason for the nucleus to stay "exactly" at the
center of the orbit when the Coulomb force in the transverse plane is
totally ignored.

In contrast, in the $z$ direction, we have more complicated situation,
namely, the magnetic field leads to the effective interaction potential
$C(z)$, which is finite at $z=0$ and is Coulomb-like in the long-range
asymptotic. Hence, the position of the nucleus remains to be fixed in
the $z$ direction relative to the orbit of the electron (bound state).

Due to the (crude) Coulomb approximation of Sect.~\ref{Cappr}, the
electron probability density of the hydrogen atom under very intense
magnetic field, at $n=1$, appears to be of rather nontrivial form
schematically depicted in Fig.~\ref{Fig:double}.

More accurate approach may yield some different electron charge
distributions but we expect that the symmetry requirements (namely, the
axial symmetry and the (anti)symmetry under inversion $z\to-z$) would
lead again to a toroidal orbit, which is \textsl{splitted into two}
identical coaxial orbits (the torii are separated) similar to that
shown in Fig.~\ref{Fig:double}, or to an \textsl{unsplitted} one (the
two torii are superimposed) as shown in Fig.~\ref{Fig:single}.

\begin{figure}[ht]
\begin{center} \epsfxsize=6cm
\parbox{\epsfxsize}{\epsffile{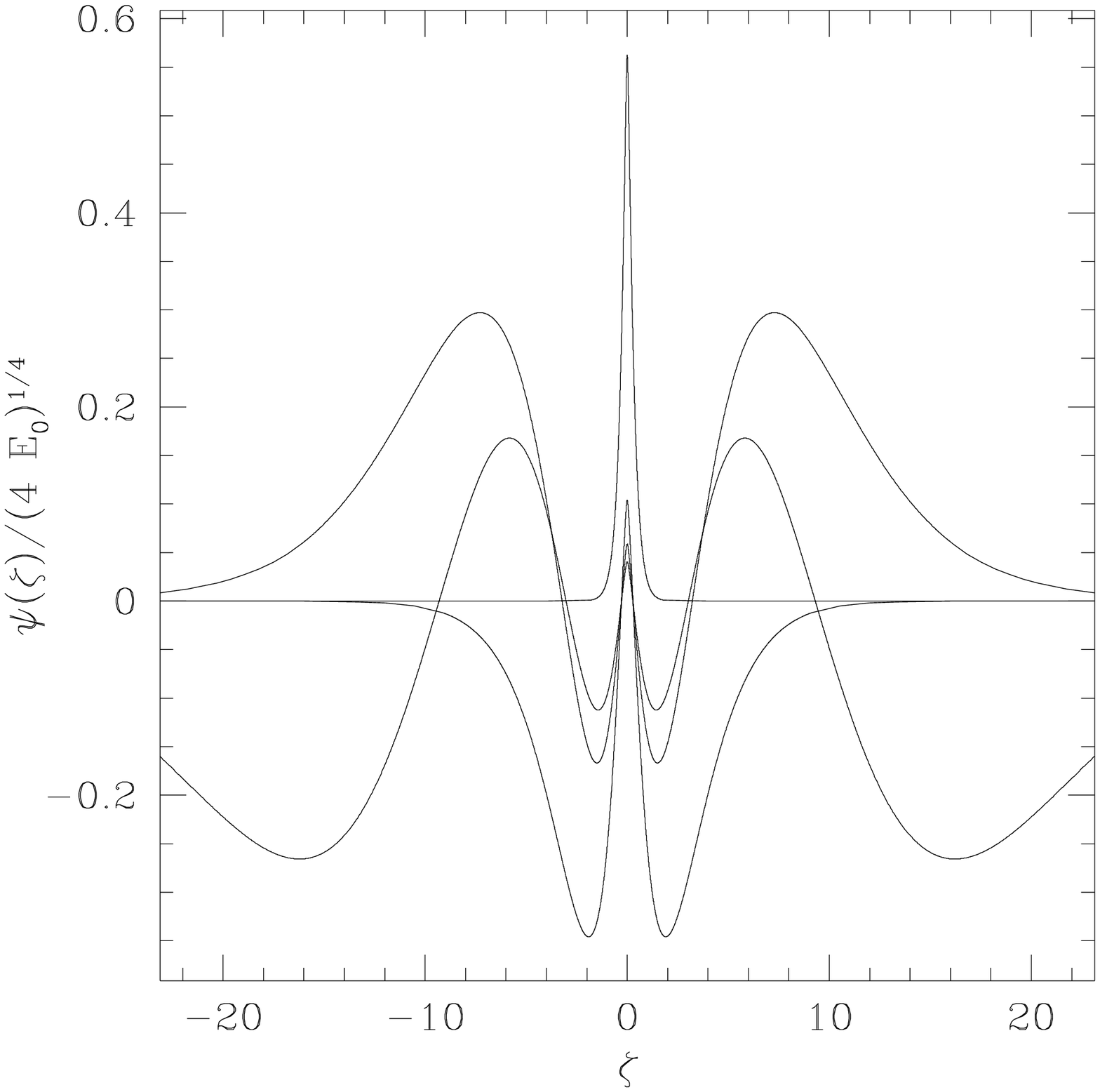}}
\parbox{\epsfxsize}{\epsffile{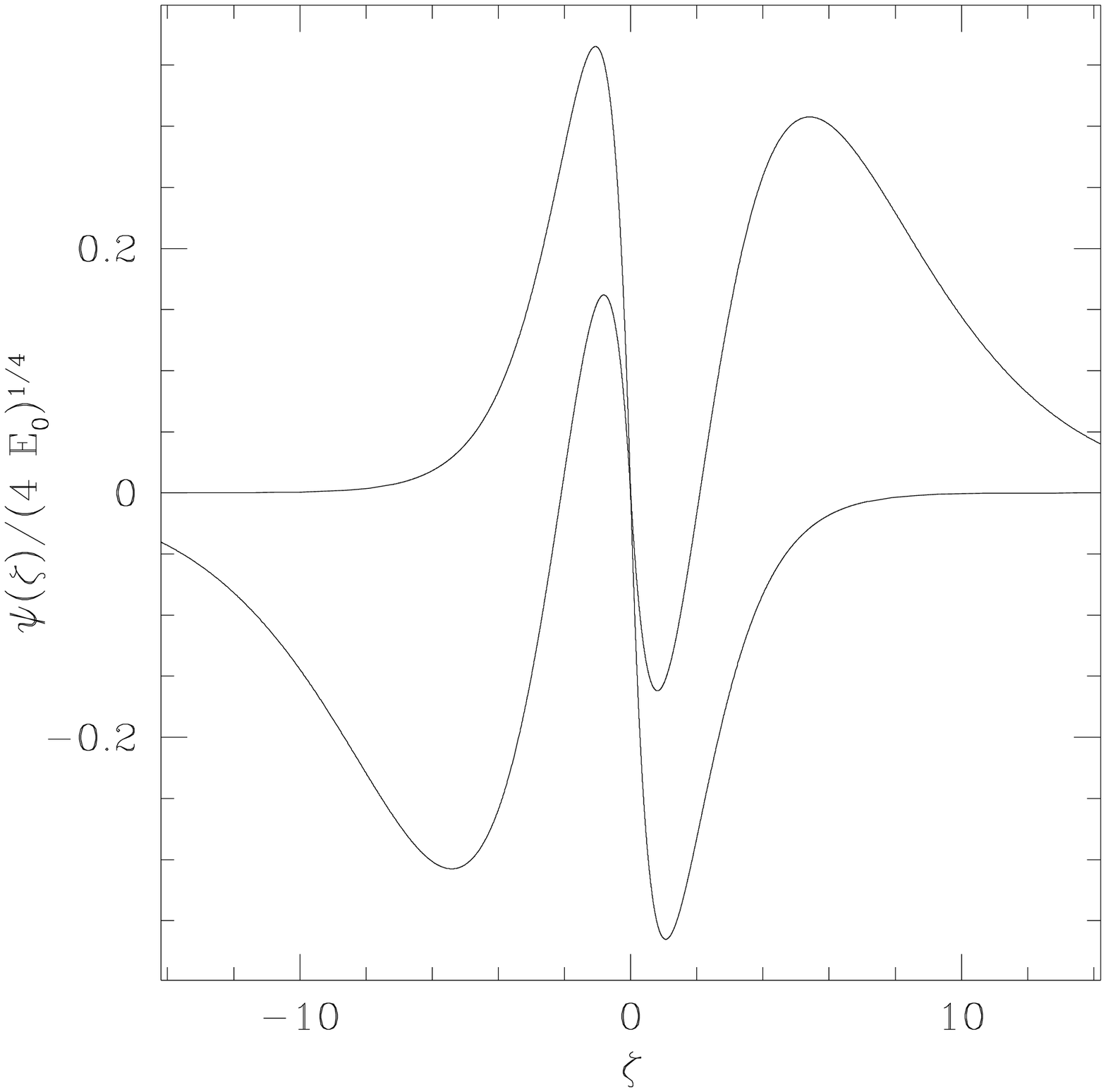}}
\end{center}
\caption{The axial wavefunctions of hydrogen in an intense magnetic
field (analytic calculation) for $B=4.7\cdot 10^{12}$ Gauss. The first
four even states with axial excitations, $|000\rangle$ (ground state),
$|002\rangle$, $|004\rangle$, and $|006\rangle$ (left panel), and odd
states $|001\rangle$ and $|003\rangle$  (right panel) are depicted.
Here,  $B=4.7\cdot 10^{12}$ Gauss, ground state energy is $E=-15.58$
Rydberg, $n=1/\sqrt{15.58}$, $\zeta=2z/n$ corresponds to $x$ in our
notation; $z$ in a.u., 1 a.u. = $0.53\cdot 10^{-8}$ cm (reproduction of
Figure~3 by Heyl and Hernquist \cite{Heyl}).} \label{Fig:Heyl3}
\end{figure}

Accurate analytic calculation of the ground and excited hydrogen wave
functions made by Heyl and Hernquist \cite{Heyl} in the adiabatic
approximation leads to the longitudinal parts of the wave functions
(the vertical $z$ dependence) shown in Fig.~\ref{Fig:Heyl3}, which
reproduces the original Fig.~3 of their work; $\zeta=2\pi\alpha
z/\lambda_e$; $B = 4.7\cdot10^{12}$ Gauss. They used the modified
Coulomb potential of the type (\ref{Cmod}), and the additional set of
linearly independent solutions of the one-dimensional modified Coulomb
problem in the form
\be\label{2solution}
(|x|+x_m)e^{-(|x|+x_m)/2} {_1}\!F_1(1-n, 2, |x|+x_m)
\int^{|x|+x_m}\frac{e^t}{(t\, {_1}\!F_1(1-n, 2, t))^2}\, dt,
\ee
where $m=0$ corresponds to the ground state. For the ground state with
$n=1/\sqrt{15.58}$, i.e. the binding energy is 15.58 Rydberg, they
found $x_0 = 0.141$, which corresponds to $B = 4.7\cdot 10^{12}$ Gauss.
This result is in agreement with the study made in
Sect.~\ref{Cmodappr}.

\begin{figure}[ht]
\begin{center} \epsfxsize=6cm
\parbox{\epsfxsize}{\epsffile{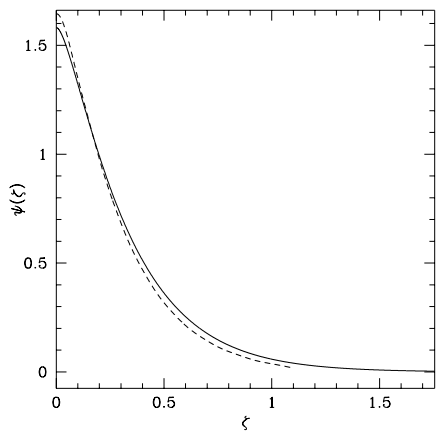}}
\end{center}
\caption{The axial ground state $|000\rangle$ wavefunction of hydrogen
in an intense magnetic field, analytic calculation (solid curve) and
numerical result (dashed line); $B=4.7\cdot 10^{12}$ Gauss, $E=-15.58$
Rydberg, $n=1/\sqrt{15.58}$, $\zeta=2z/n$ corresponds to $x$ in our
notation; $z$ in a.u., 1 a.u. = $0.53\cdot 10^{-8}$ cm (reproduction of
Figure~4 by Heyl and Hernquist \cite{Heyl}).} \label{Fig:Heyl4}
\end{figure}

One can see from Fig.~\ref{Fig:Heyl3} that the peak of the ground state
wave function $|000\rangle$ is at the point $z=0$ (see also
Fig.~\ref{Fig:Heyl4} for a close view) while the largest peaks of the
excited wave functions are away from the point $z=0$ (as it was
expected to be). Consequently, the associated longitudinal probability
distributions (square modules of the wave functions multiplied by the
volume factor of the chosen coordinate system) are symmetric with
respect to $z\to-z$, and their maxima are placed in the center $z=0$
for the ground state, and away from the center for the excited states.

In contrast to the double toroidal orbit (Fig.~\ref{Fig:double})
implied by the Coulomb approximation of Sect.~\ref{Cappr}, the modified
Coulomb potential approach suggests a \textsl{single} toroidal orbit
(Fig.~\ref{Fig:single}) for the \textsl{ground} state charge
distribution of the hydrogen atom. This confirms Santilli's study
\cite{Santilli} of the hydrogen atom in strong magnetic field. The
excited states $|00\nu\rangle$ are characterized by a double toroidal
orbit schematically depicted in Fig.~\ref{Fig:double}.

The computed ground state $|000\rangle$ binding energy of the hydrogen
atom for different field intensities are \cite{Heyl}:

\smallskip
\begin{center}
\begin{tabular}{r|c}
\hline
Magnetic field $B$ \vrule height 14pt width 0pt  & Binding energy, $|000\rangle$ state\\
(Gauss)              & (Rydberg)\\
\hline
$4.7 \times 10^{12}$\vrule height 14pt width 0pt &15.58\\
$9.4 \times 10^{12}$ &18.80\\
$23.5 \times 10^{12}$&23.81\\
$4.7 \times 10^{13}$ &28.22\\
$9.4 \times 10^{13}$ &33.21\\
$23.5 \times 10^{13}$&40.75\\
$4.7 \times 10^{14}$ &47.20\\
\hline
\end{tabular}
\end{center}
\smallskip

They calculated first-order perturbative corrections to the above
energies and obtained the values, which are in a good agreement with
the results by Lai \cite{Lai} (see Fig.~\ref{Fig:Lai1}) and Ruder {\it
et al.} \cite{Sokolov}.

It is remarkable to note that in very strong magnetic field, the energy
differences between the excited states and the ground state of the
hydrogen atom are small in comparison to the absolute value of the
ground state energy. The excited states are represented by the double
torii configurations since the peaks of the associated wave functions
spread far away from the center $z=0$ in both $+z$ and $-z$ directions.

The above ground state electron charge distributions strongly support
the study made by Santilli \cite{Santilli} who proposed the polarized
toroidal electron orbit in the hydrogen atom under the action of strong
magnetic field as a physical picture at the foundation of the new
chemical species of magnecules. The only addition to this picture is
that the the excited states are characterized by the double torii
configuration.

Since zero field ground state case is characterized by perfect
spherically symmetric electron charge distribution of the hydrogen
atom, intermediate intensities of the magnetic field are naturally
expected to imply a distorted spherical distribution. However, deeper
analysis is required for the intermediate magnetic field intensities
because the adiabatic approximation is not longer valid in this case.
Namely, an interesting problem is to solve the Schr\"odinger equation
(\ref{Schr}) for the case when intensity of the magnetic field is not
very strong, $B \leq B_0$, with the corresponding characteristic Landau
ground state energy of about the Bohr energy. In this case, both the
Coulomb and magnetic interactions should be taken at the same footing
that leads to complications in its analytic study.

\begin{figure}[ht]
\begin{center} \epsfxsize=9cm
\parbox{\epsfxsize}{\epsffile{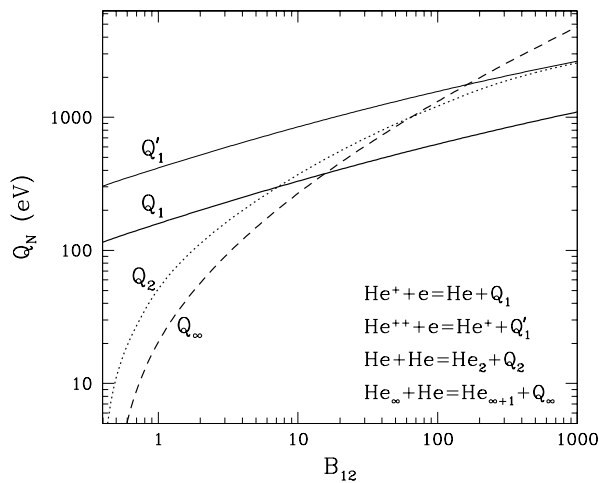}}
\end{center}
\vspace{-2cm}\caption{Ionization energies of He atom exposed to very
strong magnetic field (numerical solutions); $B_{12} = B\cdot 10^{-12}$
Gauss (reproduction of Fig.~5 by Lai~\cite{Lai}).} \label{Fig:Lai5}
\end{figure}

As to the multi-electron atoms, an interesting problem is to study
action of a very strong external magnetic field on He atom (see. e.g.,
Heyl and Hernquist \cite{Heyl} and  Lai \cite{Lai};
Fig.~\ref{Fig:Lai5}) and on the multi-electron heavy atoms, with outer
electrons characterized by a \textsl{nonspherical} charge distribution,
such as the $p$-electrons in Carbon atom, orbitals of which penetrate
the orbitals of inner electrons. Very intense magnetic field would
force such outer electrons to follow \textsl{small round} Landau
orbits. In addition to the effect of a direct action of the magnetic
field on the inner electrons, a series of essential rearrangements of
the whole electron structure of the atom seems to be occur with a
variation of the external magnetic field strength. Indeed, the magnetic
field competes with both the Coulomb energy, which is different for
different states of electrons, and the electron-electron interactions,
including spin pairings. However, it is evident that at sufficiently
strong fields, all the electron spins are aligned antiparallel to the
magnetic field --- fully spin polarized configuration --- while at
lower field intensities various partial spin polarized configurations
are possible.

\begin{figure}[ht]
\begin{center} \epsfxsize=9cm
\parbox{\epsfxsize}{\epsffile{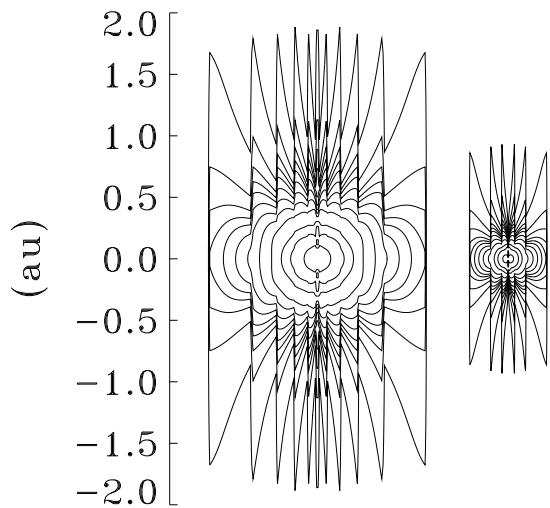}}
\end{center}
\caption{Contour plots of the $(r,z)$ plane electron density of the
iron atom according to density matrix theory at two different magnetic
field strengths, $10^{11}$ Gauss (left) and $10^{12}$ Gauss (right).
The outermost contour encloses 99\% of the negative charge, the next
90\%, then 80\% etc., and the two innermost 5\% and 1\% respectively
(reproduction of Fig.~5 by Johnsen and Yngvason \cite{Johnsen}).}
\label{Fig:Johnsen1}
\end{figure}

In accord to the numerical calculations based on the density matrix
theory approach by Johnsen and Yngvason \cite{Johnsen}, which is in
quite good agreement with the result of Hartree-Fock approach at very
strong magnetic field intensities, the inner domain in the iron atom
(26 electrons) is characterized by slightly distorted spherically
symmetric distribution even at the intensities as high as $B=100B_0
\dots 1000B_0$. The outer domain appears to be of specific, highly
elongated distribution along the direction of the magnetic field as
shown in Fig.~\ref{Fig:Johnsen1}. The possible interpretation that the
inner electrons remain to have spherical distribution while outer
electrons undergo the squeeze seems to be not correct unless the spin
state of the iron atom is verified to be partially polarized. So, we
can conclude that all the electrons are in highly magnetically
polarized state (Landau state mixed a little by Coulomb interaction),
and the electron structure is a kind of \textsl{Landau multi-electron
cylindrical shell}, with spins of all the electrons being aligned
antiparallel to the magnetic field (fully spin polarized
configuration).

Another remark regarding Fig.~\ref{Fig:Johnsen1} is that the contours
indicating nearly spherical distribution will always appear since the
Coulomb center (nucleus) is not totally eliminated from the
consideration (non-adiabatic approximation), and it forces a spherical
distribution to some degree, which evidently depends on the distance
from the center (closer to the center, more sphericity). We note that
Fig.~\ref{Fig:Johnsen1} is in qualitative agreement with
Fig.~\ref{Fig:double} in the sense that the predicted charge
distribution reveals symmetry under the inversion $z\to-z$, with the
characteristic $z$-elongated toroidal orbits.

An exciting problem is to study H$_2$ molecule under the action of
strong external static uniform magnetic field using Schr\"odinger
equation. Since the binding energy of H$_2$ molecule is of about 4.75
eV (much less than 13.6 eV of individual hydrogen atoms) and two
interacting electrons of spin 1/2 participate dynamics we expect
interesting physical predictions. For example, it is interesting to
analyze the problem of bonding, in the presence of a strong magnetic
field.

\begin{figure}[ht]
\begin{center} \epsfxsize=6cm
\parbox{\epsfxsize}{\epsffile{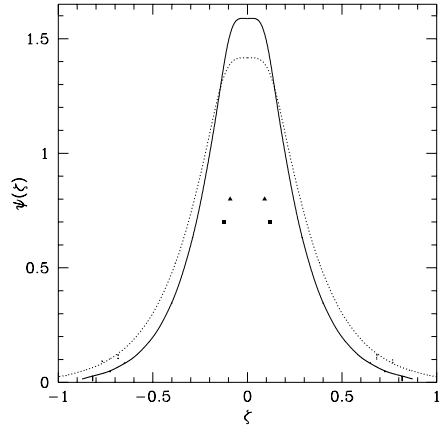}}
\end{center}
\caption{The ground and first-excited state of H$_2^+$ ion.  The solid
line traces $|000\rangle$, and the dashed line follows
$|0\,$-$1\,0\rangle$. The triangles give the positions of the protons
for the ground state and the squares for the excited state. Magnetic
field $B=4.7\cdot 10^{12}$ Gauss is pointed along the internuclear
axis; $\zeta=2\pi\alpha z/\lambda_e$ denotes $z$ in a.u.; 1 a.u. =
$0.53\cdot10^{-8}$ cm (reproduction of Figure~5 by Heyl and Hernquist
\cite{Heyl}).} \label{Fig:Heyl5}
\end{figure}

However, before studying H$_2$ molecule it would be very useful to
investigate much simpler two-center system, $H_2^+$ ion, as it can give
valuable information on the features of a two-center system under the
action of a strong magnetic field. We refer the interested reader to
Refs.~\cite{Lai, Heyl} for details of studies on H$_2^+$ ion and H$_2$
molecule in strong magnetic field. Figure~\ref{Fig:Heyl5} displays
ground and first excited state wave functions of $H_2^+$ ion calculated
by Heyl and Hernquist \cite{Heyl}.

Finally, it should be noted that studies on the atomic and molecular
systems in strong magnetic fields and related issues are becoming
extensive last years. We refer the reader to a recent review by Lai
\cite{Lai} for an extensive list of references, and below present
summary of some selected recent papers.

The work by Kravchenko {\it et al.} \cite{Kravchenko} uses an
orthogonal approach and achieves very high precision by solving the
general problem of the hydrogen atom in an arbitrarily strong magnetic
field. This analytical work is of much importance as it provides solid
ground to analyze few-electron and multi-electron atoms, and simple
diatomic molecules.

A series of papers by Schmelcher {\it et al.} \cite{Schmelcher} is
devoted to a systematic study of various simple atomic, ionic and
molecular systems in strong magnetic fields via Hartree-Fock method.
One of the papers by Schmelcher {\it et al.} considers the carbon atom.

Analytical approximations have been constructed  by Potekhin
\cite{Potekhin} for binding energies, quantum mechanical sizes and
oscillator strengths of main radiative transitions of hydrogen atoms
arbitrarily \textsl{moving} in magnetic fields $\simeq 10^{12} \ldots
10^{13}$~Gauss. This approach is of much importance in investigating an
interaction of highly magnetized hydrogen atoms in a gaseous phase.

Computational finite element method for \textsl{time-dependent}
Schr\"odinger equation has been developed by Watanabe and Tsukada
\cite{Watanabe} to study dynamics of electrons in a magnetic field that
can be used effectively for mesoscopic systems, such as small clusters.

\section*{Acknowledgements}

The author is much grateful to R.M. Santilli for discussions and
support. Additional thanks are due to M.I. Mazhitov, Karaganda State
University, for stimulating discussions.

\newpage
\appendix
\section*{Appendix}

\begin{figure}[!ht]
\begin{center} \epsfxsize=5cm
\parbox{\epsfxsize}{\epsffile{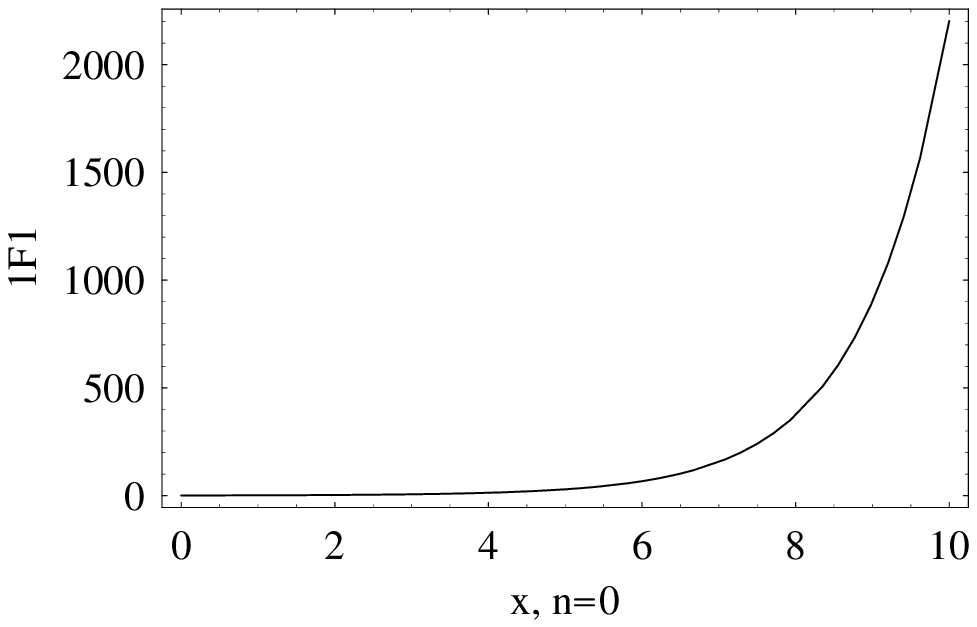}}
\parbox{\epsfxsize}{\epsffile{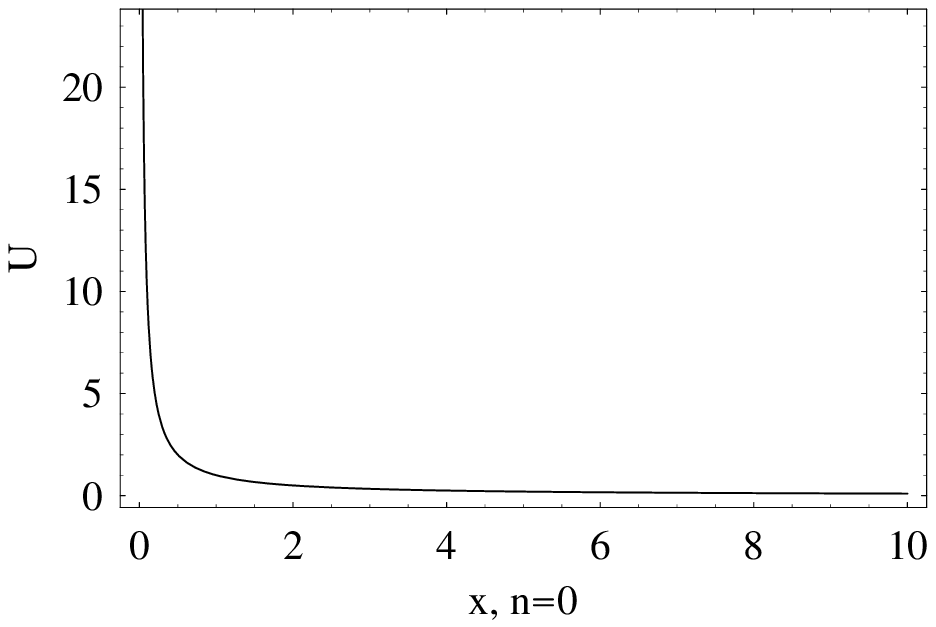}}
\end{center}
\begin{center} \epsfxsize=5cm
\parbox{\epsfxsize}{\epsffile{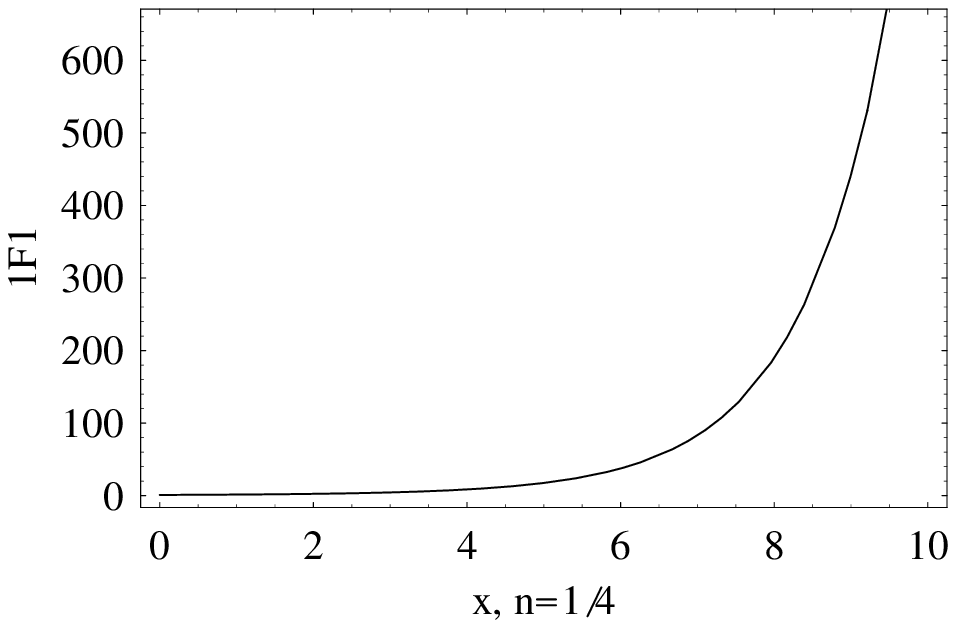}}
\parbox{\epsfxsize}{\epsffile{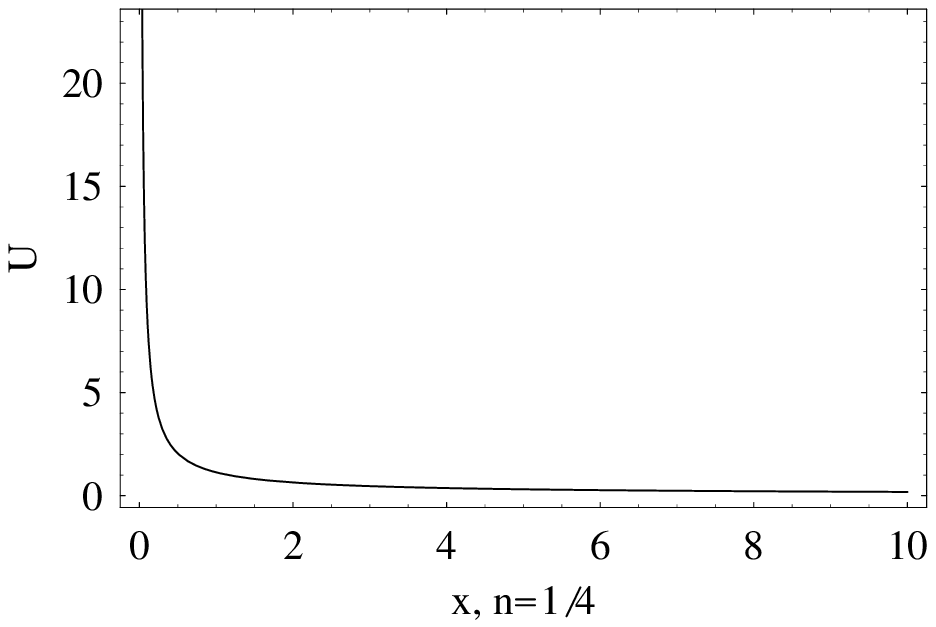}}
\end{center}
\begin{center} \epsfxsize=5cm
\parbox{\epsfxsize}{\epsffile{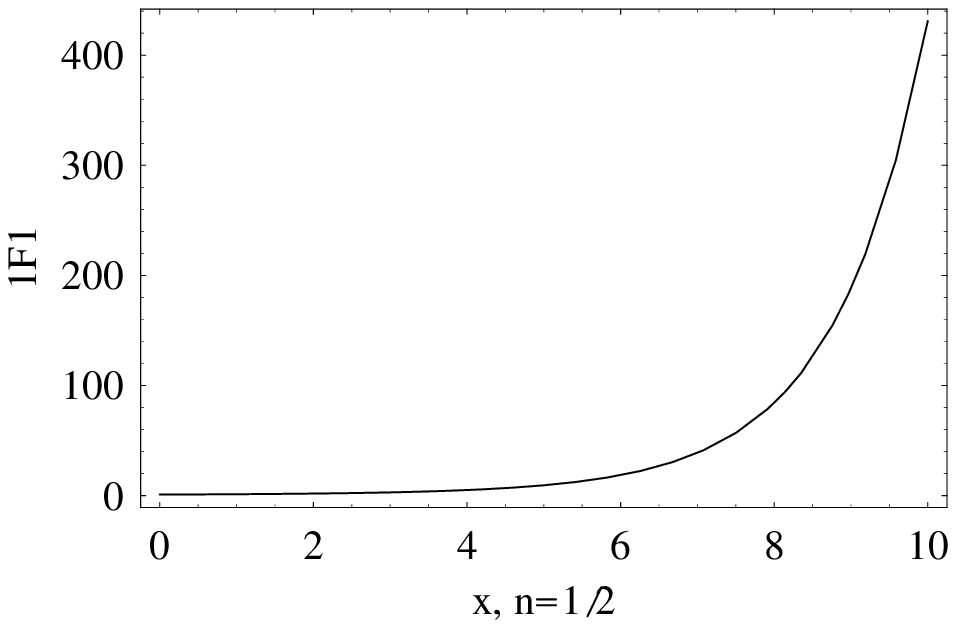}}
\parbox{\epsfxsize}{\epsffile{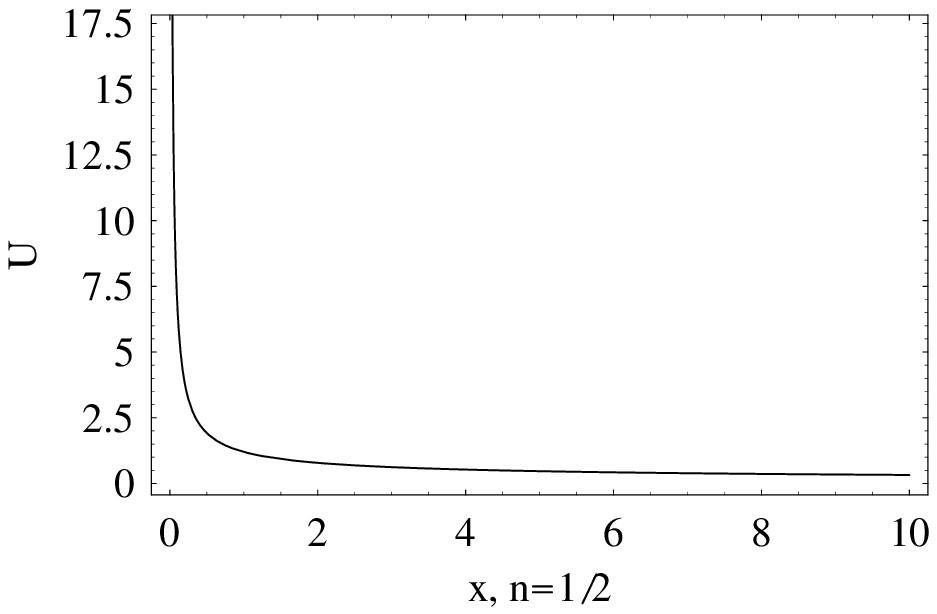}}
\end{center}
\begin{center} \epsfxsize=5cm
\parbox{\epsfxsize}{\epsffile{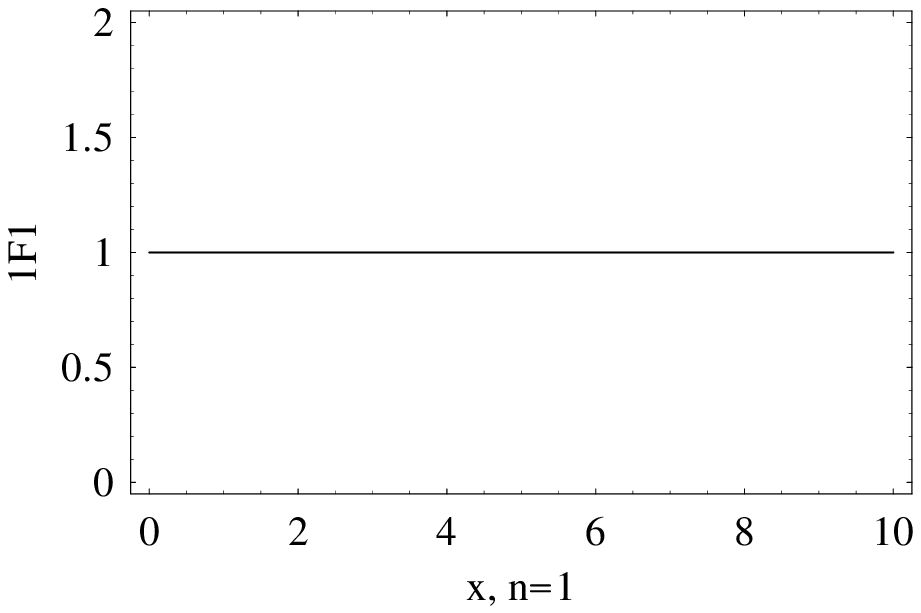}}
\parbox{\epsfxsize}{\epsffile{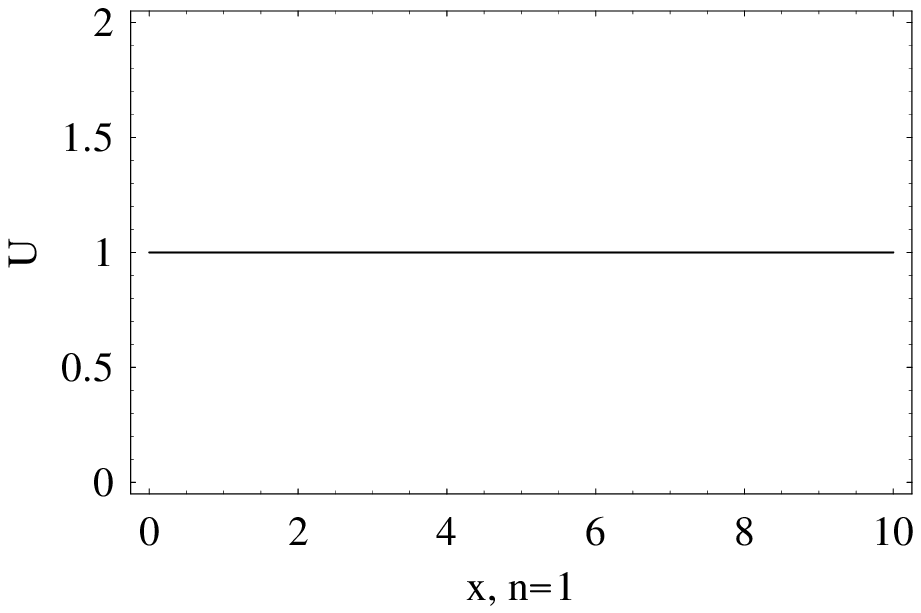}}
\end{center}
\caption{The confluent hypergeometric functions ${_1}\!F_{1}(1-n, 2,x)$
and $U(1-n,2,x)$ for various $n$.} \label{Fig:uf1}
\end{figure}

\begin{figure}[ht]
\begin{center} \epsfxsize=5cm
\parbox{\epsfxsize}{\epsffile{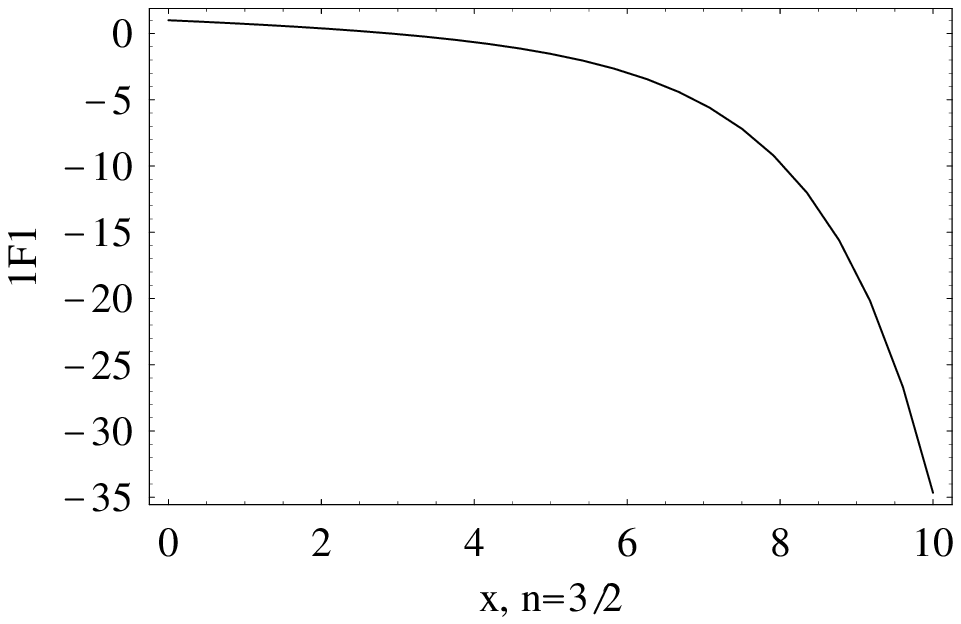}}
\parbox{\epsfxsize}{\epsffile{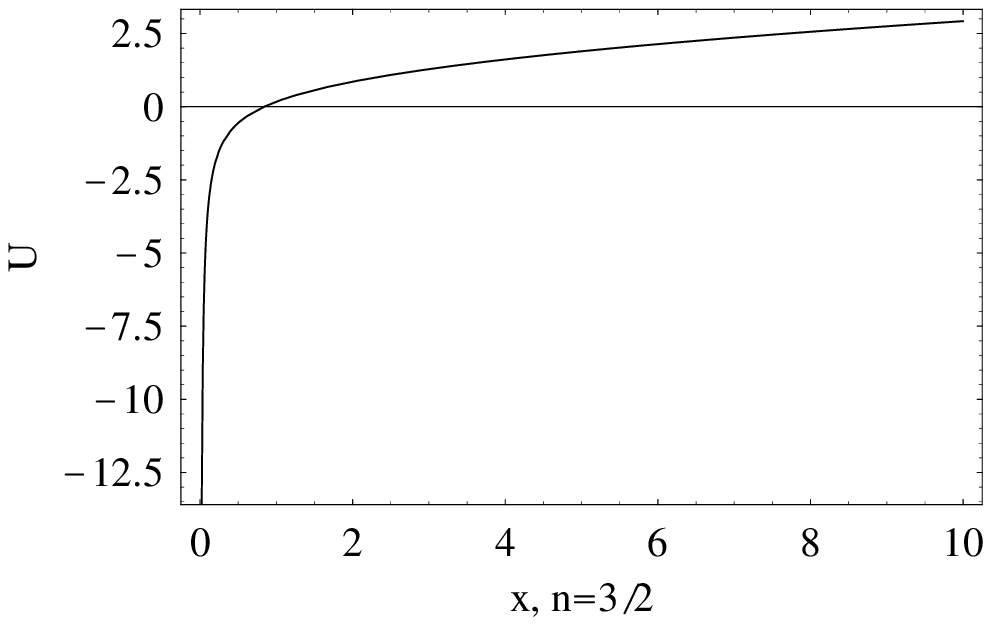}}
\end{center}
\begin{center} \epsfxsize=5cm
\parbox{\epsfxsize}{\epsffile{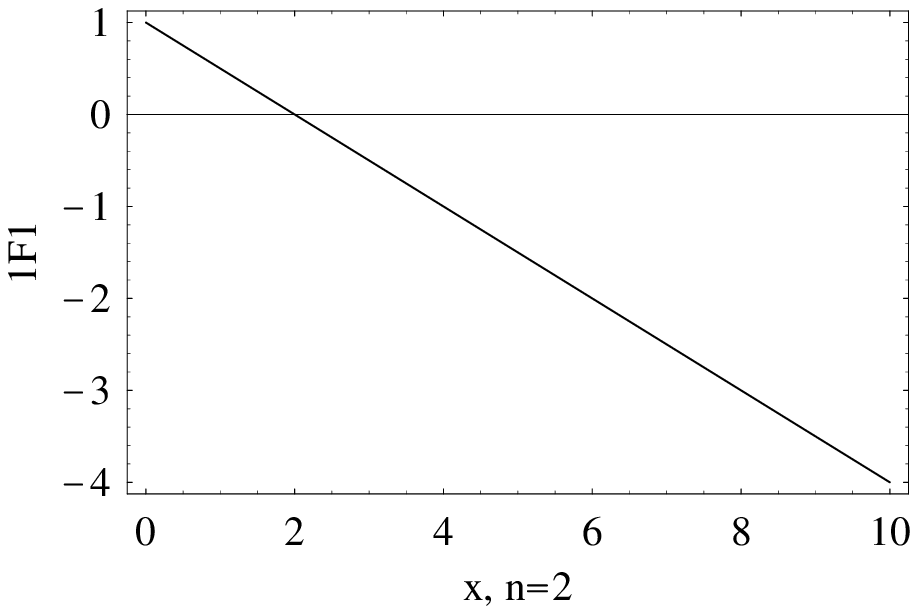}}
\parbox{\epsfxsize}{\epsffile{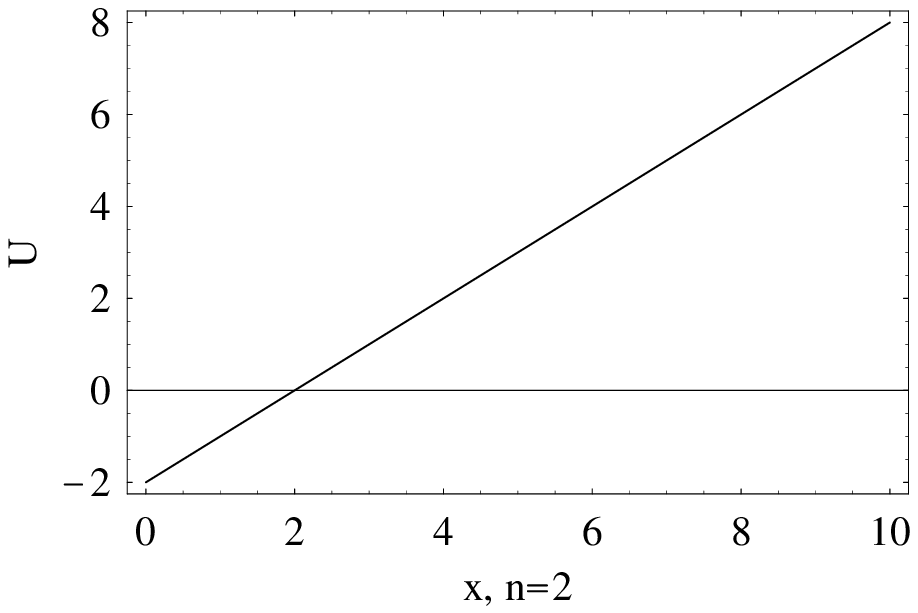}}
\end{center}
\begin{center} \epsfxsize=5cm
\parbox{\epsfxsize}{\epsffile{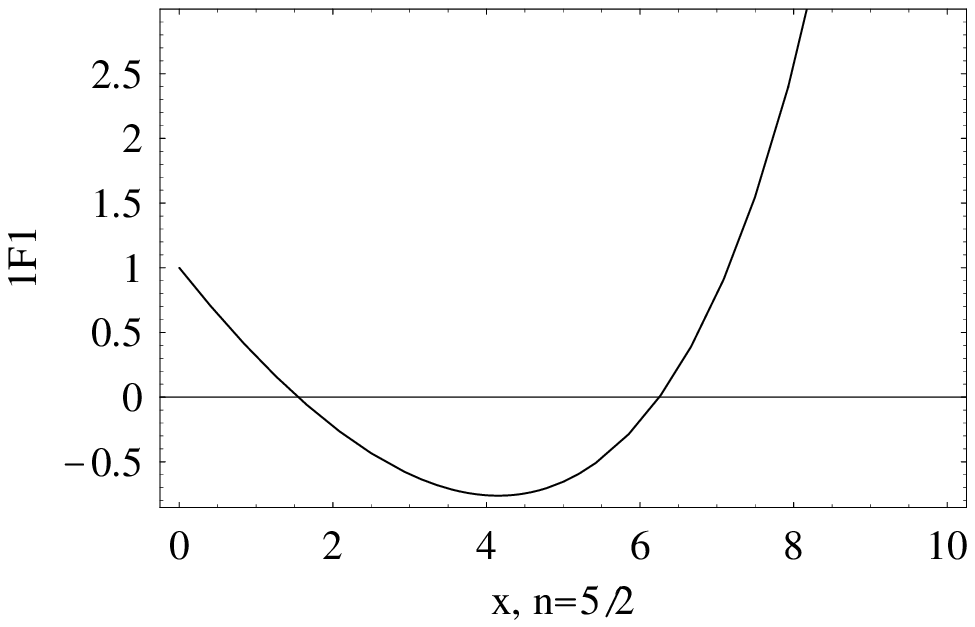}}
\parbox{\epsfxsize}{\epsffile{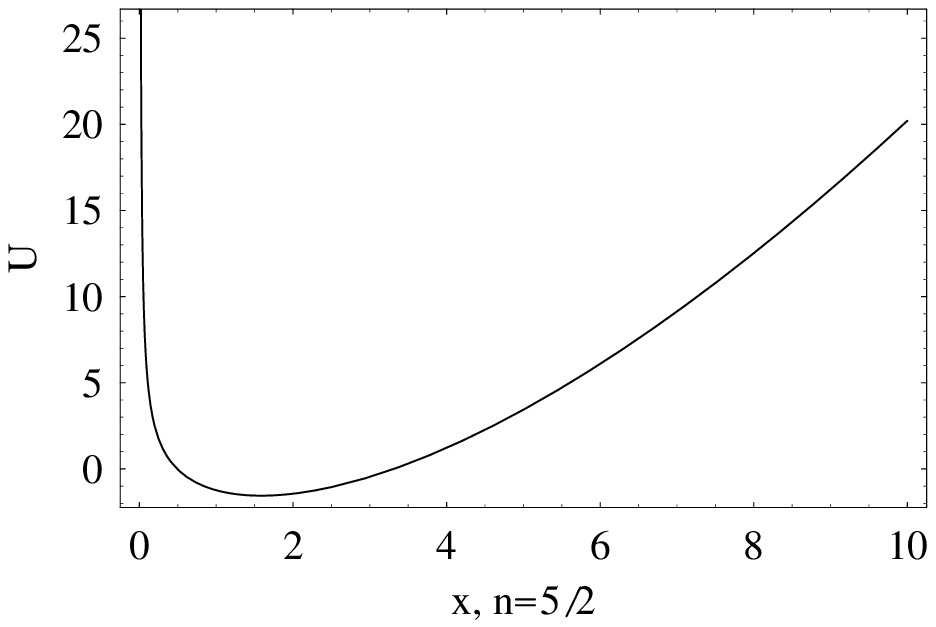}}
\end{center}
\begin{center} \epsfxsize=5cm
\parbox{\epsfxsize}{\epsffile{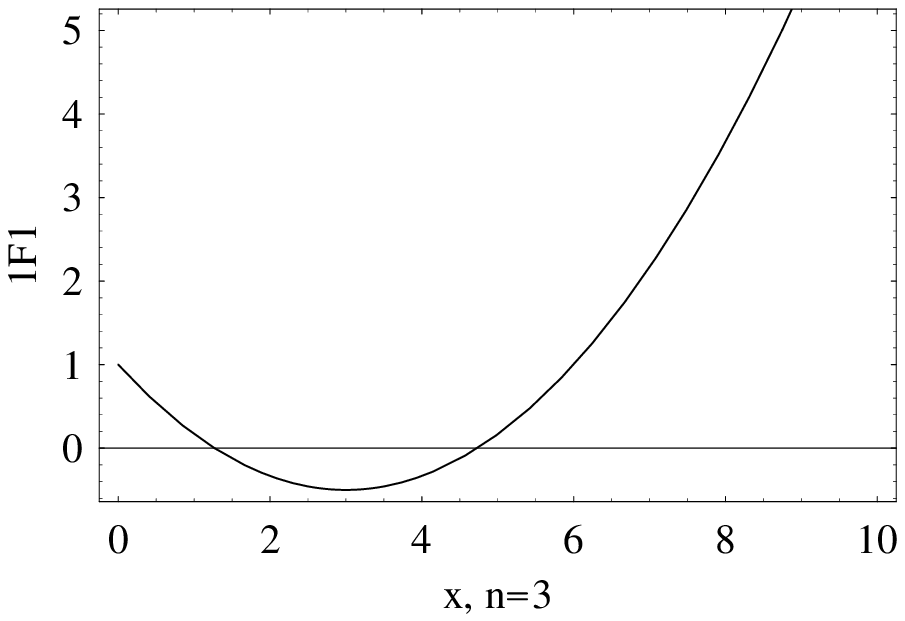}}
\parbox{\epsfxsize}{\epsffile{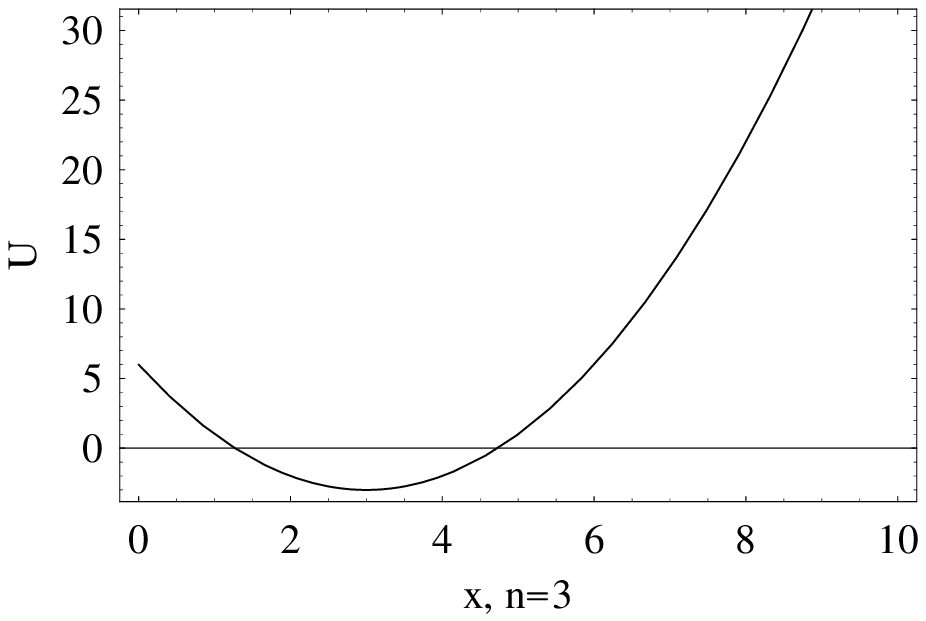}}
\end{center}
\caption{The confluent hypergeometric functions ${_1}\!F_{1}(1-n, 2,x)$
and $U(1-n,2,x)$ for various $n$.} \label{Fig:uf2}
\end{figure}

\end{document}